\documentclass[pre,twocolumn,showkeys,showpacs, superscriptaddress,nofootinbib]{revtex4-1}
\usepackage{graphics,graphicx,latexsym,amssymb,amsmath,amsbsy,color}
\usepackage{hyperref}

\begin{document}
\title{Anomalous Stress Fluctuations in Athermal Two Dimensional Amorphous Solids}
\author{Yegang Wu}
\affiliation{Department of Physics and Astronomy, University of Rochester, Rochester, NY 14627, USA}
\author{Kamran Karimi}
\affiliation{Universit{\'e} Grenoble Alpes, LiPhy, F-38000 Grenoble, France}
\author{Craig~E.~Maloney}
\affiliation{Department of Mechanical and Industrial Engineering, Northeastern University, Boston,	 Massachusetts 02115, USA}
\author{S. Teitel}
\affiliation{Department of Physics and Astronomy, University of Rochester, Rochester, NY 14627, USA}
\date{\today}
\begin{abstract}
We numerically study the local stress distribution within athermal, isotropically stressed, mechanically stable, packings of bidisperse frictionless disks above the jamming transition in two dimensions.  Considering the Fourier transform of the local stress, we find evidence for algebraically increasing fluctuations in both  isotropic and anisotropic components of the stress tensor at small wavenumbers, contrary to recent theoretical predictions.  Such increasing fluctuations imply a lack of self-averaging of the stress on large length scales.
The crossover to these increasing fluctuations defines a length scale $\ell_0$, however it appears that $\ell_0$ does not vary much with packing fraction $\phi$, nor does $\ell_0$ seem to be diverging as $\phi$ approaches the jamming $\phi_J$.   We also find similar large length scale fluctuations of stress in the inherent states of a quenched Lennard-Jones liquid, leading us to speculate that such fluctuations may be a general property of amorphous solids in two dimensions.
\end{abstract}
\maketitle

%%%%%%%%%%%%%%%%%%%%%%%%%%%%%%%%%%%%%%%%%%%%%%%%%%%%%%%%%%%%%%%%%%%%%%%%%%%%%

\maketitle

\section{Introduction}

Amorphous solids abound in nature, from dense granular packings, to foams, to metallic glasses.
Amorphous solids may be considered a unique state of matter.  They have a finite shear modulus and resist shear flow, like familiar crystalline solids.  However, unlike crystalline solids, the particles are in seemingly random positions, reminding one of a liquid.  It is therefore of interest to study the properties of amorphous solids to see in which ways they might be more similar to a crystal, or to a liquid, or be uniquely different from either.

One quantity of practical importance  is the distribution of stress throughout the system.  For a crystalline solid, the stress fields vary periodically with the periodic positioning of the particles.   For a liquid, or other random particle patterns, one expects that the stress fields may vary randomly, but that the system will be self-averaging, i.e., the relative fluctuation in the total stress should decrease inversely proportionally to the square root of the averaging volume.  
In this work we consider numerically the fluctuations of stress in simple two dimensional (2D) amorphous solids.  Our focus will be on dense, athermal, mechanically stable packings of a bidisperse distribution of soft-core frictionless disks, above the jamming transition \cite{OHern}.  However we will also consider the stress distribution in the inherent structures of a quenched bidisperse Lennard-Jones liquid.  

A field theoretic model for isotropically compressed athermal 2D granular systems was proposed by Henkes and Chakraborty (HC) \cite{Henkes2}.  They argued that the fluctuations of pressure at finite wavevector $\mathbf{q}$ approach a constant as $|\mathbf{q}|\to 0$, 
\begin{equation}
\langle |\delta p_\mathbf{q}|^2\rangle = \dfrac{1}{A_0+A_2q^2+A_4q^4+\dots},
\label{pHC}
\end{equation}
and that the length scale determined by the coefficient $A_2/A_0$ remains small and finite even as the jamming transition is approached.
This result would imply spatially short ranged pressure correlations, consistent with the notion of self-averaging.  For fluctuations of the simple shear stress, their model predicts, 
\begin {equation}
\langle | (\sigma_{xy})_\mathbf{q}|^2\rangle = \dfrac{q_x^2q_y^2}{q^4}\dfrac{1}{(C_0+C_2q^2+C_4q^4+\dots)},
\label{sigHC}
\end{equation}
which results \cite{Henkes2} in power law spatial correlations that decay as $1/r^2$. However these spatial correlations are anti-correlated (i.e. negative) in the directions $\pm\mathbf{\hat x}$ and $\pm\mathbf{\hat y}$, but positively correlated in the directions $\pm\mathbf{\hat x}\pm\mathbf{\hat y}$.  When averaging over the direction of $\mathbf{r}$, we believe that the cancellation of positive and negative terms in this correlation will result in angular averaged spatial correlations that are short ranged, and [as we will argue following Eq.~(\ref{eDC})] would result in a fluctuation of shear stress that is self-averaging.

Numerical simulations \cite{Lois,WuTeitel} and experiments \cite{Lois} on granular disks have reported results consistent with these predictions by HC.
Other recent work has considered the stress correlations in the inherent structures of supercooled liquids.  
Lema{\^i}tre argued \cite{Lemaitre} that the stress field in such inherent structures should arise from a succession of spatially uncorrelated and isotropically oriented Eshelby transformations, each with an associated long-range-correlated stress field.
Chowdhury et. al. \cite{Harrowell} made a similar argumentâ that stresses arise from spatially uncorrelated and isotropically distributed force dipoles.  Both arrive at the conclusion that
spatial correlations in the shear stress field should
decay as $1/r^2$ in 2D, in agreement with HC \cite{Henkes2}.

Recently, however, Karimi and Maloney \cite{Karimi}, using simulations of much larger 2D systems then previously studied, 
considered the fluctuations of the anisotropic (deviatoric) part of the local stress tensor in soft-core disk packings.  Averaging over a window of length $R$, they found that the relative fluctuations in the average local deviatoric stress  decayed as $1/R$ for small $R$ (i.e. as the inverse square root of the volume, as expected for short-range correlated stress), but decayed more slowly at larger $R$, thus implying the presence of stress correlations on long length scales.  The crossover between these small and large $R$ behaviors was found to occur at a length scale  larger than was accessible in previous simulations and experiments on smaller systems \cite{Lois,WuTeitel}.

In this work we reexamine the fluctuations of the local stress tensor in 2D isotropically stressed, mechanically stable, packings of bidisperse frictionless disks.  Using large systems with up to $N=65536$ particles, we find that above a certain large length scale $\ell_0$, both isotropic and anisotropic components of the stress tensor show anomalously large fluctuations, consistent with the findings of Karimi and Maloney \cite{Karimi} for the anisotropic part.  We investigate how this behavior varies as the packing fraction decreases towards the jamming transition, and find that the length $\ell_0$ appears to approach a finite large constant, rather than diverging as one of the divergent length scales associated with the jamming transition.  A similar behavior has recently been observed for fluctuations of the local packing fraction \cite{WuHyper}.  We then investigate stress fluctuations in the inherent states of a Lennard-Jones interacting system, and find similar behavior as in the granular packings.  We thus speculate that anomalous stress fluctuations may be a characteristic feature of 2D amorphous solids in general, and that, contrary to the above theoretical predictions, fluctuations of the stress are not self-averaging.

Our paper is organized as follows.  In Sec.~\ref{smodel} we introduce our model for athermal, bidisperse, soft-core interacting frictionless disks in two dimensions and discuss our protocol for creating mechanically stable packings of these disks above the jamming transition.  In Sec.~\ref{sresults} we present our numerical results.  In Sec.~\ref{ssoftq} we consider the wavevector-dependant correlations of the stress in Fourier space and show that at small wavevectors they grow as the wavevector decreases, deviating from the predicted results of HC described above.  In Sec.~\ref{ssoftR} we consider the corresponding fluctuations of the stress in real space, averaged over spatial windows of increasing radius $R$.  We show that such fluctuations behave in a manner at odds with self-averaging.  In Sec.~\ref{sprotocol2} we discuss tests we have made to see if our conclusions concerning the large length scale stress fluctuations are sensitive to the particular protocol we have chosen to construct our amorphous solid configurations.  We find that they are robust.  In Sec.~\ref{sLJ} we consider, instead of soft-core disks, the inherent states of a quenched bidisperse Lennard-Jones liquid.  We find, for the wavevector-dependent stress correlations, the same anomalous behavior at small wavevectors  that we find for soft-core disks.  Finally in Sec.~\ref{sscaling} we test if the wavector-dependent stress correlations for soft-core disks scale with any of the diverging correlation lengths that have been associated with the jamming transition.  We find that they do not.  In Sec.~\ref{sconclusions} we summarize our conclusions.  In our Appendix A we discuss the accuracy of our method and provide further details concerning one of the stress correlations that is expected to vanish at long wavelengths.  In Appendix B we derive a relation between stress fluctuations at finite waver vectors $\mathbf{q}$, and fluctuations averaged over a spatial window of radius $R$.

\section{Model}
\label{smodel}

\subsection{Soft-core disks}

The main model we will consider in this work is that of athermal soft-core frictionless disks in mechanically stable equilibrium, at finite pressure above the jamming transition in two dimensions.
Our model is one that has been much studied in the literature \cite{OHern}.  We use a bidisperse distribution of $N$ circular disks with equal numbers of big and small particles with diameter ratio $d_b/d_s=1.4$.  Particles interact only when they overlap, in which case they repel with a harmonic elastic interaction, 
\begin{equation}
{\cal V}({r}_{ij})=\frac{1}{2}k_e (1-{r}_{ij}/d_{ij})^2,
\end{equation} 
where ${r}_{ij}=|\mathbf{r}_i-\mathbf{r}_j|$ is the center-to-center distance between  disks $i$ and $j$,  $d_{ij}=(d_i+d_j)/2$ the average diameter of the two disks, and $k_e$ is the coupling that sets the energy scale.  We will measure length in units such that $d_s=1$ and energy in units such that $k_e=1$.

For a system of $N$ particles at positions $\{\mathbf{r}_i\}$, the stress tensor $\boldsymbol{\Sigma}_i$ on particle $i$ is \cite{Henkes3}
\begin{equation}
\boldsymbol{\Sigma}_i=\sum_j\mathbf{s}_{ij}\otimes\mathbf{F}_{ij},\quad\mathbf{F}_{ij}=-\partial {\cal V}(r_{ij})/\partial\mathbf{r}_j,
\end{equation}
where the sum is over all particles $j$ in contact with $i$, $\mathbf{s}_{ij}$ is the displacement from the center of particle $i$ to its point of contact with particle $j$, and $\mathbf{F}_{ij}$ is the force on $j$ due to its contact with $i$.  The total stress tensor $\boldsymbol{\Sigma}$ for the entire system, and the pressure $p$, are then given by,
\begin{equation}
\boldsymbol{\Sigma}=\sum_i\boldsymbol{\Sigma_i}, \quad p=\frac{1}{2V}\mathrm{Tr}[\boldsymbol{\Sigma}],
\end{equation}
with $V$ the total system volume.
In this work we will consider primarily mechanically stable packings with {\em isotropic} total system stress, 
\begin{equation}
\boldsymbol{\Sigma}=\Gamma\mathbf{I},\qquad\Gamma = pV,
\end{equation}
with $\mathbf{I}$ the identity tensor.  

\subsection{Preparation protocol}
\label{sprotocol}

\begin{figure}[h]
\begin{center}
\includegraphics[width=1.8in]{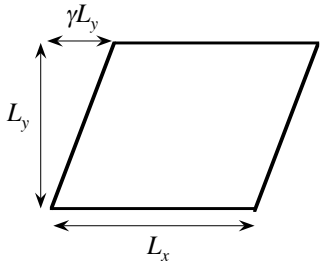}
\caption{Geometry of our system box.  $L_x$ and $L_y$ are the lengths in the $\mathbf{\hat x}$ and $\mathbf{\hat y}$ directions, and $\gamma$ is the skew ratio.  Lees-Edwards boundary conditions are used.
}
\label{box}
\end{center}
\end{figure}

To prepare such isotropically stressed packings, we use the following procedure \cite{WuHyper}.
Our system box, into which our $N$ particles are placed, is characterized by three parameters, $L_x,L_y,\gamma$, as illustrated in Fig.~\ref{box}. $L_x $ and $L_y$ are the lengths of the box in the $\mathbf{\hat x}$ and $\mathbf{\hat y}$ directions, while $\gamma$ is the skew ratio of the box.  We use Lees-Edwards boundary conditions \cite{LeesEdwards} to periodically repeat this box throughout all space.

We introduce a modified energy function $\tilde U$ that depends on the particle positions $\{\mathbf{r}_i\}$, as well as the box parameters $L_x,L_y,\gamma$,
\begin{equation}
\tilde U= U+\Gamma(\ln L_x +\ln L_y),\quad U=\sum_{i<j}{\cal V}_{ij}(r_{ij}).
\label{eUtilde}
\end{equation}
Here $\Gamma=pV$ is the target value for the total system isotropic stress.
The interaction energy $U$ depends implicitly on the box parameters $L_x,L_y,\gamma$ via the boundary conditions, and one can show that,
\begin{equation}
\begin{aligned}
L_x\frac{\partial U}{\partial L_x}=-\Sigma_{xx}+\gamma\Sigma_{xy},&\quad\frac{\partial U}{\partial \gamma}=-\Sigma_{xy},\\ 
L_y\frac{\partial U}{\partial L_y}=-\Sigma_{yy}-\gamma\Sigma_{xy}.&
\end{aligned}
\label{ebox}
\end{equation}

Starting from an initial configuration, we then minimize $\tilde U$
with respect to both particle positions and box parameters.  Minimizing with respect to particle positions $\{\mathbf{r}_i\}$ results in a vanishing net force on each particle.  Minimizing with respect to the box parameters $L_x,L_y,\gamma$ results, via Eqs.~(\ref{ebox}), in the desired isotropic total stress tensor,
\begin{equation}
\Sigma_{xx}=\Sigma_{yy}=\Gamma, \quad \Sigma_{xy}=0 \quad (\Sigma_{xy}=\Sigma_{yx}). 
\end{equation}
Further details of our algorithm may be found in Ref.~\cite{WuHyper}.  A discussion of the accuracy of our method is given in Appendix A.  

For our initial starting configurations, we use a square box with $L_x=L_y=L$, $\gamma=0$, and place particles down completely at random, with $L$ chosen to give an initial packing fraction,
\begin{equation}
\phi_\mathrm{init}=
\dfrac{\pi N}{2L^2}\left[\left(\dfrac{d_s}{2}\right)^2+\left(\dfrac{d_b}{2}\right)^2\right].
\label{ephinit}
\end{equation}
Unless otherwise stated, we take $\phi_\mathrm{init}=0.84$, slightly below the jamming transition.
Our results at each value of $\Gamma$ are averaged over 1000--10000 (depending on the system size) independently generated isotropic configurations.  Configurations are generated independently at each value of $\Gamma$.

It will be convenient to parametrize our configurations by the intensive stress per particle, 
\begin{equation}
\tilde p= \Gamma/N = p V/N.
\end{equation}  
We will consider a range of $\tilde p=0.00014$ -- 0.01831, spanning  over two orders of magnitude.  At each fixed $\tilde p$, since our protocol involves variation of the box parameters, each individual minimized configuration has a slightly different box area $L_xL_y$, and so a slightly different packing fraction $\phi$.  The above range of $\tilde p$ corresponds to a range of average packing fractions $\langle\phi\rangle = 0.8416$ -- 0.8857 \cite{WuHyper}.  We will use systems with $N=8192$ -- 65536 particles. 
In the limit of an {\em infinitely} large system the jamming transition, where $\tilde p$ falls to zero upon decreasing the packing fraction, occurs at $\phi_J\approx 0.8416$ for our particular protocol \cite{WuHyper}; our finite size systems, however, will have a small but finite $\tilde p$ at this $\phi_J$ due to finite size effects.

\subsection{Stress tensor, correlations, and fluctuations}
\label{sstress}

To distinguish the isotropic vs the anisotropic parts of the stress, we decompose the 2D symmetric stress tensor $\boldsymbol{\Sigma}_i$ on particle $i$ into three scalar parameters, $\Gamma_i$, $\delta\Gamma_i$, and $\Sigma_{xyi}$,
\begin{equation}
\boldsymbol{\Sigma}_i=\Gamma_i\left[ 
\begin{array}{cc}
1&0\\0&1
\end{array}\right]
+\delta\Gamma_i\left[ 
\begin{array}{cc}
1&0\\0&-1
\end{array}\right]
+\Sigma_{xyi}\left[
\begin{array}{cc}
0&1\\1&0
\end{array}\right].
\label{edecomposition}
\end{equation}
The first piece, proportional to $\Gamma_i$, is the isotropic part that determines the pressure, $\Gamma=\sum_i\Gamma_i=pV$.  The second two pieces give the shear stress, with the deviatoric stress  $\tau_i$ given by $\tau_i^2=\delta\Gamma_i^2+\Sigma_{xyi}^2$.  Note that under a rotation of coordinates by an angle $\theta$, the stress tensor retains the same form as Eq.~(\ref{edecomposition}), but with,
\begin{equation}
\begin{aligned}
\Gamma_i^\prime&=\Gamma_i\\
\delta\Gamma_i^\prime&=\delta\Gamma_i\cos 2\theta - \Sigma_{xyi}\sin 2\theta\\
\Sigma_{xyi}^\prime&=\Sigma_{xyi}\cos 2\theta + \delta\Gamma_i\sin 2\theta,
\end{aligned}
\label{rotate}
\end{equation}
so that for $\theta=\pi/4$,  $\delta\Gamma_i\to\Sigma_{xyi}^\prime$ and $\Sigma_{xyi}\to-\delta\Gamma_i^\prime$.

To study fluctuations of stress at finite wavevectors $\mathbf{q}$ we introduce the Fourier transform,
\begin{equation}
\boldsymbol{\Sigma}_\mathbf{q}=\sum_i\mathrm{e}^{i\mathbf{q}\cdot\mathbf{r}_i}\boldsymbol{\Sigma}_i,
\end{equation}
with $\Gamma_\mathbf{q}$, $\delta\Gamma_\mathbf{q}$ and $\Sigma_{xy\mathbf{q}}$ defined similarly.  To relate our work to that of HC, we note that their  $p_\mathbf{q}$ is our $\Gamma_\mathbf{q}$, and their $\sigma_{xy\mathbf{q}}$ is our $\Sigma_{xy\mathbf{q}}$.

The allowed wavevectors consistent with the Lees-Edwards boundary conditions are 
\begin{equation}
\mathbf{q}=2\pi[(m_1/L_x)\mathbf{\hat x}+(m_2/L_y-\gamma m_1/L_x)\mathbf{\hat y}],
\label{eqs}
\end{equation} 
with $m_1$ and $m_2$ integer.  Since each configuration at a given total value of $\Gamma$ has a slightly different value of $L_x$, $L_y$ and $\gamma$, these set of allowed $\mathbf{q}$ vary slightly from configuration to configuration.  However, since $\langle L_x\rangle=\langle L_y\rangle$ and $\langle\gamma\rangle=0$, and the fluctuations about these averages are very small for our large systems sizes  (see Appendix of Ref.~\cite{WuHyper}), these differences are negligible and so when averaging stress over different configurations in our ensemble, we average the stress at wavevectors corresponding to common values of $m_1$ and $m_2$.\footnote{We have also considered a constant volume ensemble in which the set of allowed $\mathbf{q}$ are identical from sample to sample. In that case we find that the resulting stress correlations remain unchanged from what we find in our constant stress ensemble, thus indicating that no artifacts are introduced by averaging at constant $m_1$ and $m_2$. See Sec.~\ref{sprotocol2}.}

To quantify stress fluctuations at finite wavevector $\mathbf{q}$ we define the correlations,
\begin{equation}
\begin{aligned}
C_\Gamma(\mathbf{q})&=\frac{1}{V}\langle\Gamma_\mathbf{q}\Gamma_{-\mathbf{q}}\rangle\\
C_{\delta\Gamma}(\mathbf{q})&=\frac{1}{V}\langle\delta\Gamma_\mathbf{q}\delta\Gamma_{-\mathbf{q}}\rangle\\
C_{\Sigma_{xy}}(\mathbf{q})&=\frac{1}{V}\langle\Sigma_{xy\,\mathbf{q}}\Sigma_{xy\,-\mathbf{q}}\rangle,
\label{eCq}
\end{aligned}
\end{equation}
where $\langle\dots\rangle$ denotes an average over  independently generated packings.
{\em If} stress fluctuations are {\em isotropic}, then we expect from Eq.~(\ref{rotate}) that $C_\Gamma(\mathbf{q})$ will be independent of the direction of $\mathbf{q}$, and that,
\begin{equation}
C_{\delta\Gamma}(\mathbf{q})=C_{\Sigma_{xy}}(\pm \mathbf{q}^R)
\label{rotate2}
\end{equation}
where $\mathbf{q}^R$ is $\mathbf{q}$ rotated by $\pm 45^\circ$.
In this work we will consider $\mathbf{q}$ in two different directions: $m_1=0$ along the $\mathbf{\hat y}$ direction, and $m_1=m_2$, which on average is along the $\mathbf{\hat e}_{x+y}=[\mathbf{\hat x}+\mathbf{\hat y}]/\sqrt{2}$ direction.

To quantify stress fluctuations in real space, we define the measure,
\begin{equation}
\Delta_\Gamma(R)=[\langle \Gamma_R^2\rangle-\langle\Gamma_R\rangle^2]/(\pi R^2),
\label{eDR}
\end{equation}
where $\Gamma_R= \sum_{i\in R}\Gamma_i$ is the sum of stresses for all particles whose center lies within a randomly placed circular window of radius $R$.  If the $\Gamma_i$ are uncorrelated beyond some length scale $\xi\ll R$, we expect that $\Delta_\Gamma(R)$ becomes constant as $R$ increases.  We similarly define $\Delta_{\delta\Gamma}(R)$ and $\Delta_{\Sigma_{xy}}(R)$.

As we show in Appendix B, the measure of real space fluctuations $\Delta_X(R)$ ($X=\Gamma, \delta\Gamma, \Sigma_{xy}$) is related to the correlation $C_X(\mathbf{q})$ by the relation,
\begin{equation}
\Delta_X(R)=\frac{\pi R^2}{V}\sum_{\mathbf{q}\ne 0}C_X(\mathbf{q})
f^2(|\mathbf{q}|R),
\label{eDC}
\end{equation} 
where $f(u)= (2/u^2)\int_0^u dv v J_0(v)$, $J_0$ is the Bessel function of the first kind, and the sum is over all wavevectors $\mathbf{q}$ consistent with the Lees-Edwards boundary conditions excluding the point $\mathbf{q}=0$.  Taking the infinite system limit $V\to\infty$, we have $(1/V)\sum_\mathbf{q}\to (1/2\pi)^2\int d^2q$, and we get,
\begin{equation}
\Delta_X(R)=\dfrac{1}{2}\int_0^\infty d\kappa \kappa \>\bar C_X(\kappa/R)f^2(\kappa),
\label{eDC2}
\end{equation}
where $\bar C_X(q)$ is the average of $C_X(\mathbf{q})$ over the direction of $\mathbf{q}$.  Since $f^2(0)=1$, and $f^2(u)\sim u^{-3}$ for $u\gtrsim 2$ \cite{WuHyper}, when $R$ is sufficiently large, it will be the small $q$ limiting values of $\bar C_X(q)$ that determine the value of the integral.  Thus if $\bar C_X(q\to 0)$ is finite, the integral becomes independent of $R$ as $R$ gets sufficiently large, and so $\Delta_X(R)$ becomes constant and the system is self-averaging.  If the predictions of HC hold, then clearly the pressure fluctuations of Eq.~(\ref{pHC}) give $\bar C_\Gamma(q\to 0)$ is finite, and since the angular average of $q_x^2q_y^2=q^4/8$, then similarly the shear stress fluctuations of Eq.~(\ref{sigHC}) give $\bar C_{\Sigma_{xy}}(q\to 0)$ is finite; hence the predictions of HC imply that the stress should be self-averaging.

\section{Results}
\label{sresults}

\subsection{Soft-core particles: Correlations in $\mathbf{q}$-space}
\label{ssoftq}

\begin{figure}[h!]
\includegraphics[width=3.4in]{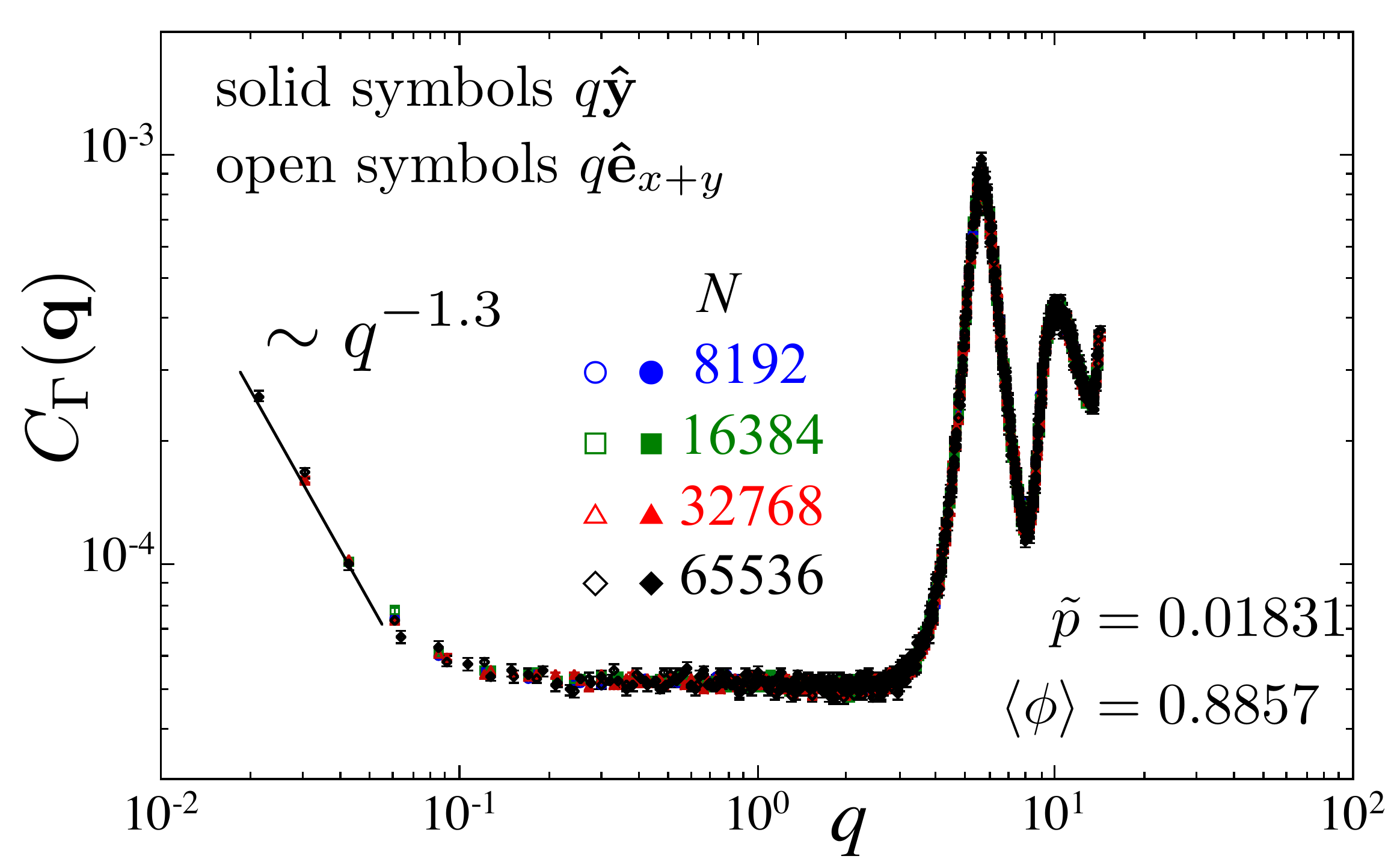}
\caption{(Color online) Fluctuation of the isotropic part of the stress $C_\Gamma(\mathbf{q})=\langle\Gamma_\mathbf{q}\Gamma_{-\mathbf{q}}\rangle/V$ vs $q$, at wavevectors $\mathbf{q}=q\mathbf{\hat y}$ (solid symbols) and $\mathbf{q}=q\mathbf{\hat e}_{x+y}$ (open symbols), at stress per particle $\tilde p=\Gamma/N=0.01831$ above jamming.  Here $\langle\phi\rangle = 0.8857$, compared to $\phi_J\approx 0.8416$.  Results are shown for systems with different number of particles $N$.  Solid line at small $q$ has slope $-1.3$.
}
\label{Gq-1200}
\end{figure}

We first consider $C_\Gamma(\mathbf{q})$, which is equivalent to the fluctuations of the pressure.  By construction, the total system stress is isotropic.  If local fluctuations are also on average isotropic, then since $\Gamma$ is the isotropic part of the stress tensor we expect that $C_\Gamma(\mathbf{q})$ depends only on $|\mathbf{q}|$ \cite{Henkes2}.  In Fig.~\ref{Gq-1200} we plot $C_\Gamma(\mathbf{q})$ vs $q$ for the two directions $q\mathbf{\hat y}$ and $q\mathbf{\hat e}_{x+y}$.  We show results for our largest total stress per particle $\tilde p =\Gamma/N=0.01831$, for several system sizes $N$.  We see that $C_\Gamma(\mathbf{q})$ is independent of $N$, and independent of the direction of $\mathbf{q}$, for the entire range of $|\mathbf{q}|$.  For a range of small $0.1\lesssim q\lesssim 1$, $C_\Gamma(\mathbf{q})$ is roughly constant, in agreement with the theory of HC \cite{Henkes2}.  However, below $q_0\approx 0.1$, $C_\Gamma(\mathbf{q})$ departs from the HC prediction, showing a roughly algebraic increase as $q$ decreases, $C_\Gamma(\mathbf{q})\sim q^{-1.3}$, though we do not have enough small $q$ data points to determine this power law with any serious accuracy.

\begin{figure}[t!]
\includegraphics[width=3.4in]{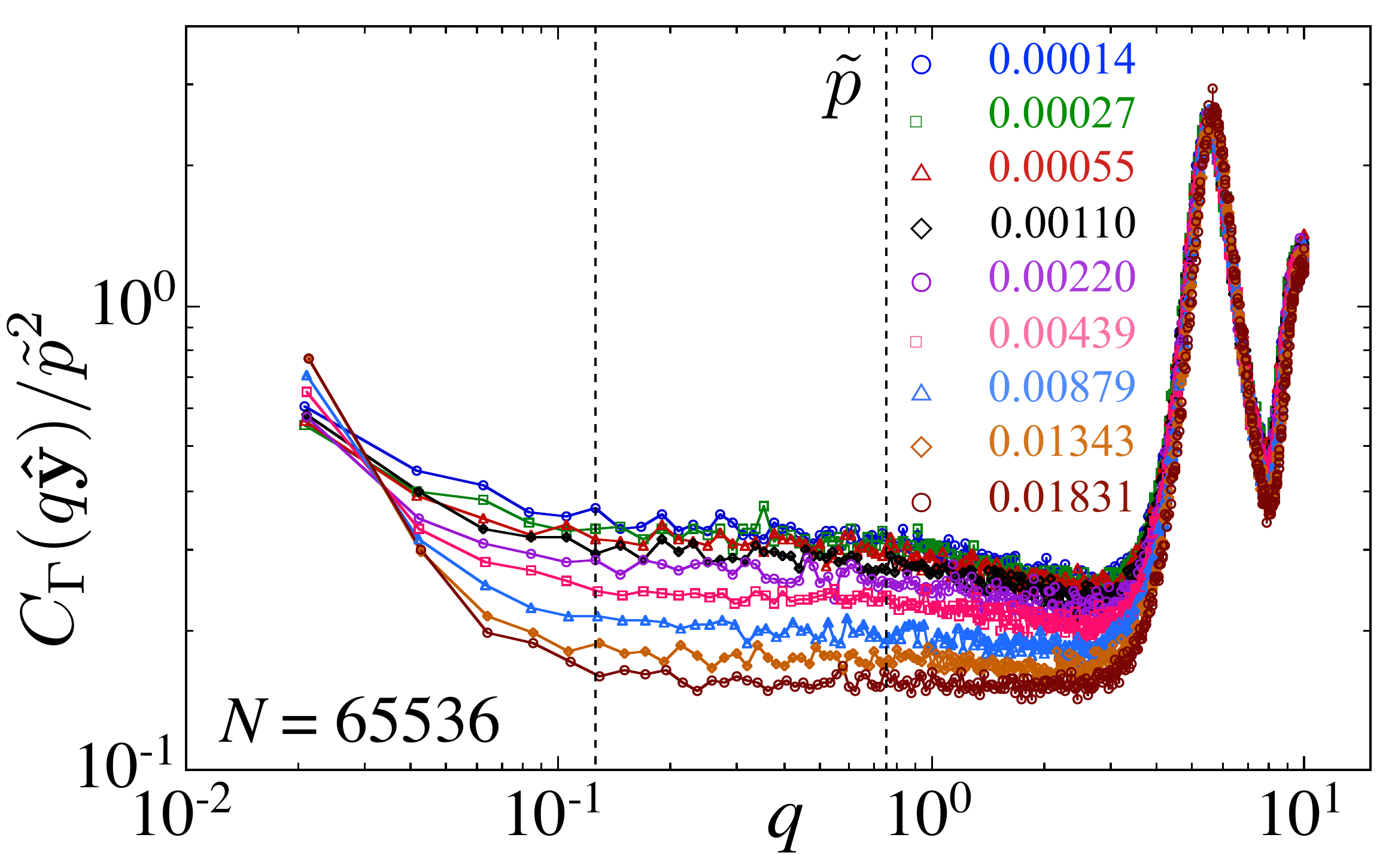}
\caption{(Color online) Fluctuation of the isotropic part of the stress  normalized by the stress per particle $\tilde p$ squared, $C_\Gamma(\mathbf{q})/\tilde p^2$, at wavevectors $\mathbf{q}=q\mathbf{\hat y}$.  Curves are for different values of $\tilde p=\Gamma/N$, going from $\tilde p= 0.00014$ on top to $0.01831$ on bottom.  System has $N=65536$ particles.  Vertical dashed lines delimit the range of $q$ where $C_\Gamma(\mathbf{q})$ is roughly constant.  
}
\label{Gqotildep2-vs-q}
\end{figure}

Next we consider $C_\Gamma(\mathbf{q})$ at other values of $\tilde p$, approaching the jamming transition.  HC have argued \cite{Henkes2} that $C_\Gamma(\mathbf{q})$ should scale proportional to the square of the stress, so in Fig.~\ref{Gqotildep2-vs-q} we plot $C_\Gamma(\mathbf{q})/\tilde p^2$  vs $q$ for $\mathbf{q}=q\mathbf{\hat y}$, for the single system size $N=65536$. 
Several features are evident in this plot.  (i) Within a range of small wavevector $0.1\lesssim q \lesssim 1$ we see that $C_\Gamma(\mathbf{q})/\tilde p^2$ is roughly constant, as found in Fig.~\ref{Gq-1200}. (ii) Within this range, the curves appear to be approaching a common value as $\tilde p$ decreases, consistent with the $\tilde p^2$ scaling of HC.  (iii) As $q$ increases above $\sim 1$, the fluctuations start to decrease as $q$ increases; this crossover, indicated by the right most vertical dashed line, is consistent with the earlier results of HC and defined their ``$\xi$".  As HC found, we see that this $\xi$ shows little variation with $\tilde p$ for the range of $\tilde p$ shown here. (iv) As $q$ decreases below $q_0\sim 0.1$, fluctuations increase roughly algebraically.  As $\tilde p$ decreases, the exponent of this power law (i.e. the slope of the plotted curves) appears to decrease.  This crossover $q_0$, roughly indicated by the left most vertical dashed line, decreases somewhat, but does not appear to be vanishing, as $\tilde p$ decreases, and so the corresponding length scale $\ell_0\approx 2\pi/q_0\approx 60$ appears to remain finite even as the jamming transition $\tilde p\to 0$ is approached.

\begin{figure}[h!]
\includegraphics[width=3.4in]{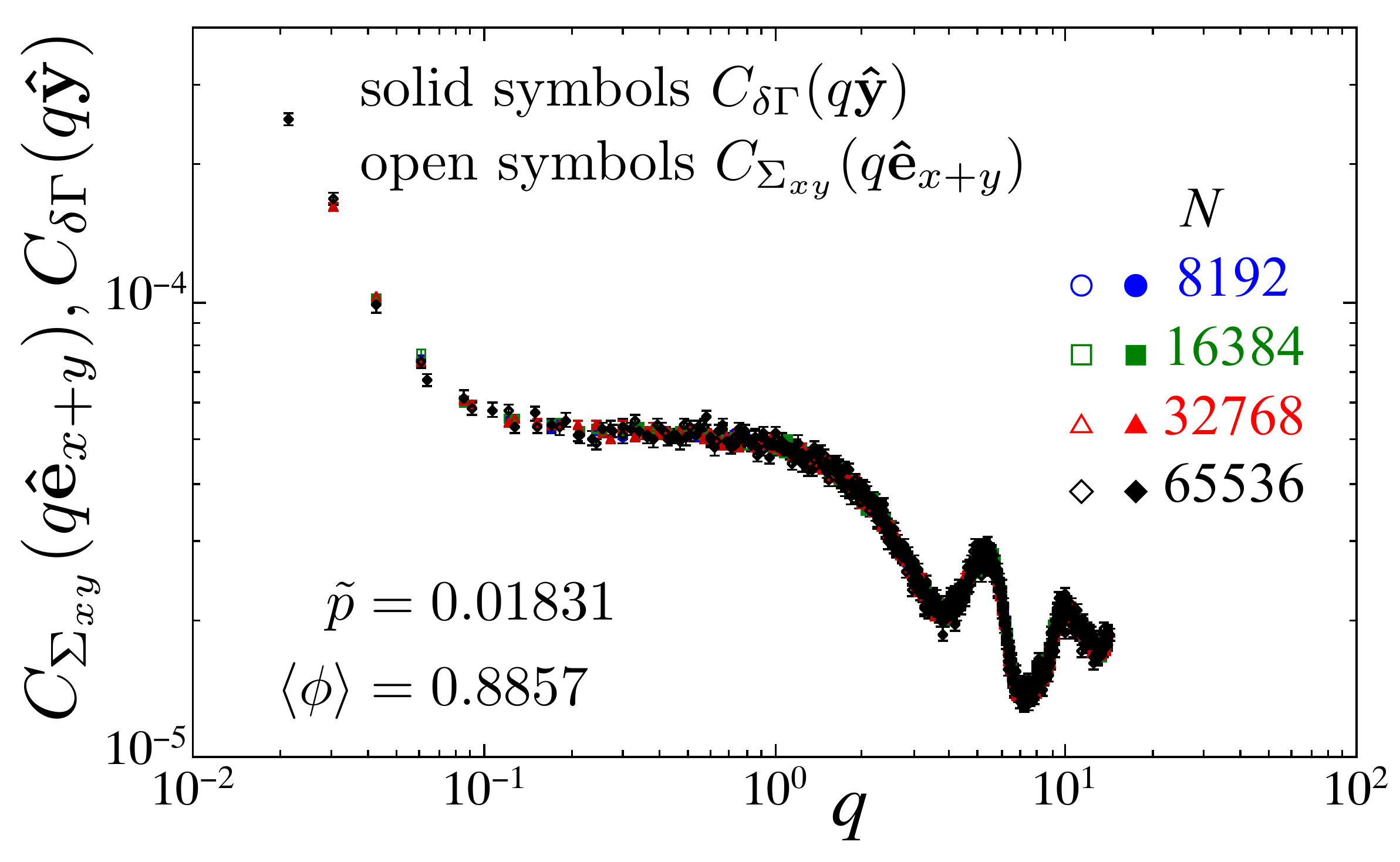}
\caption{(Color online) Fluctuation of the anisotropic part of the stress $C_{\Sigma_{xy}}(\mathbf{q})=\langle\Sigma_{xy\mathbf{q}}\Sigma_{xy-\mathbf{q}}\rangle/V$ at wavevectors $\mathbf{q}=q\mathbf{\hat e}_{x+y}$ (open symbols), and $C_{\delta\Gamma}(\mathbf{q})=\langle\delta\Gamma_\mathbf{q}\delta\Gamma_{-\mathbf{q}}\rangle/V$ at wavevectors $\mathbf{q}=q\mathbf{\hat y}$ (solid symbols).  
Stress per particle is $\tilde p=\Gamma/N=0.01831$ above jamming.  Results are shown for systems with different number of particles $N$.  
}
\label{dGq-1200}
\end{figure}

\begin{figure}[h!]
\includegraphics[width=3.4in]{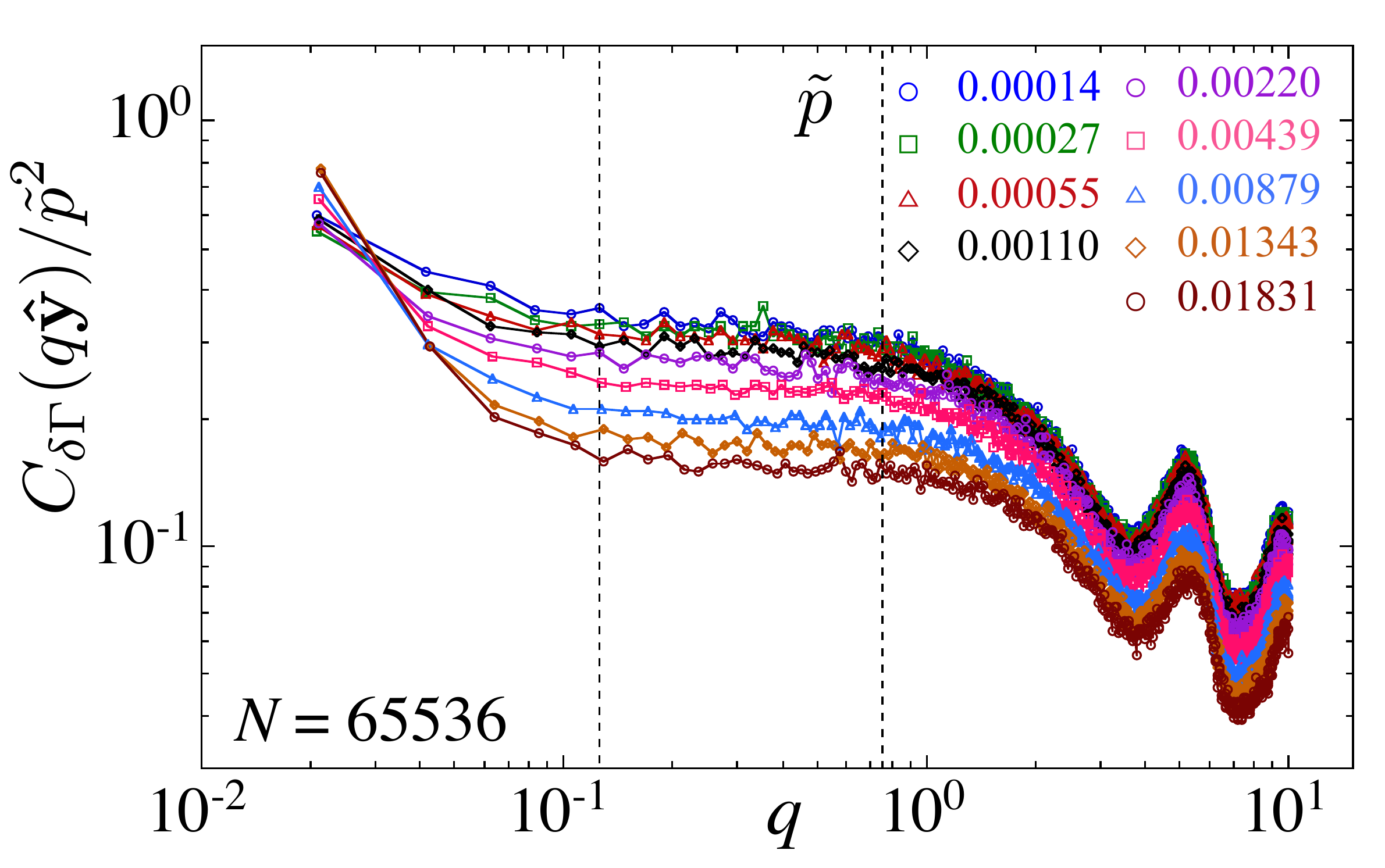}
\caption{(Color online) Fluctuation of the anisotropic part of the stress  normalized by the stress per particle $\tilde p$ squared, $C_{\delta\Gamma}(q\mathbf{\hat y})/\tilde p^2$, vs $q$.  Curves are for different values of $\tilde p=\Gamma/N$, going from $\tilde p= 0.00014$ on top to $0.01831$ on bottom.  System has $N=65536$ particles.  Vertical dashed lines delimit the range of $q$ where $C_{\delta\Gamma}(q\mathbf{\hat y})$ is roughly constant.  
}
\label{dGqotildep2-vs-q}
\end{figure}

Next we consider the anisotropic part of the stress tensor, corresponding to the shear stress. According to Eq.~(\ref{rotate2}), if fluctuations are isotropic, we expect that  $C_{\delta\Gamma}(q\mathbf{\mathbf{\hat y}})=C_{\Sigma_{xy}}(q\mathbf{\hat e}_{x+y})$.
In Fig.~\ref{dGq-1200} we therefore plot these two correlations vs $q$ for different system sizes $N$, for total stress per particle $\tilde p=0.01831$.  We see no dependence on $N$, and we see the agreement of the two correlations as expected.  
From the prediction of Eq.~(\ref{sigHC}) by HC, 
the small-$q$ behavior of $C_{\Sigma_{xy}}(\mathbf{q})\sim q_x^2q_y^2/q^4$.  Thus for $\mathbf{q}$ in direction $\mathbf{\hat e}_{x+y}$, where $q_x=q_y$, we expect $C_{\Sigma_{xy}}(q\mathbf{\hat e}_{x+y})\to\mathrm{constant}$ as $q\to 0$.  In contrast we find that, while $C_{\Sigma_{xy}}(q\mathbf{\hat e}_{x+y})$ is roughly constant over a range of small $0.1\lesssim q\lesssim 1$, it suddenly increases as $q$ decreases to small values, similar to the behavior found in Fig.~\ref{Gq-1200} for $C_\Gamma(\mathbf{q})$.  In Fig.~\ref{dGqotildep2-vs-q} we consider the correlation $C_{\delta\Gamma}(q\mathbf{\hat y})$ at different values of $\tilde p$ for the system of size $N=65536$, plotting $C_{\delta\Gamma}(q\mathbf{\hat y})/\tilde p^2$ vs $q$.  As in Fig. ~\ref{Gqotildep2-vs-q}, we find that as $\tilde p$ decreases, the curves appear to approach a common limiting curve and the boundaries of the flat region at small $q$  (dashed vertical lines) do not appear to vary much with $\tilde p$.

\begin{figure}
\includegraphics[width=3.4in]{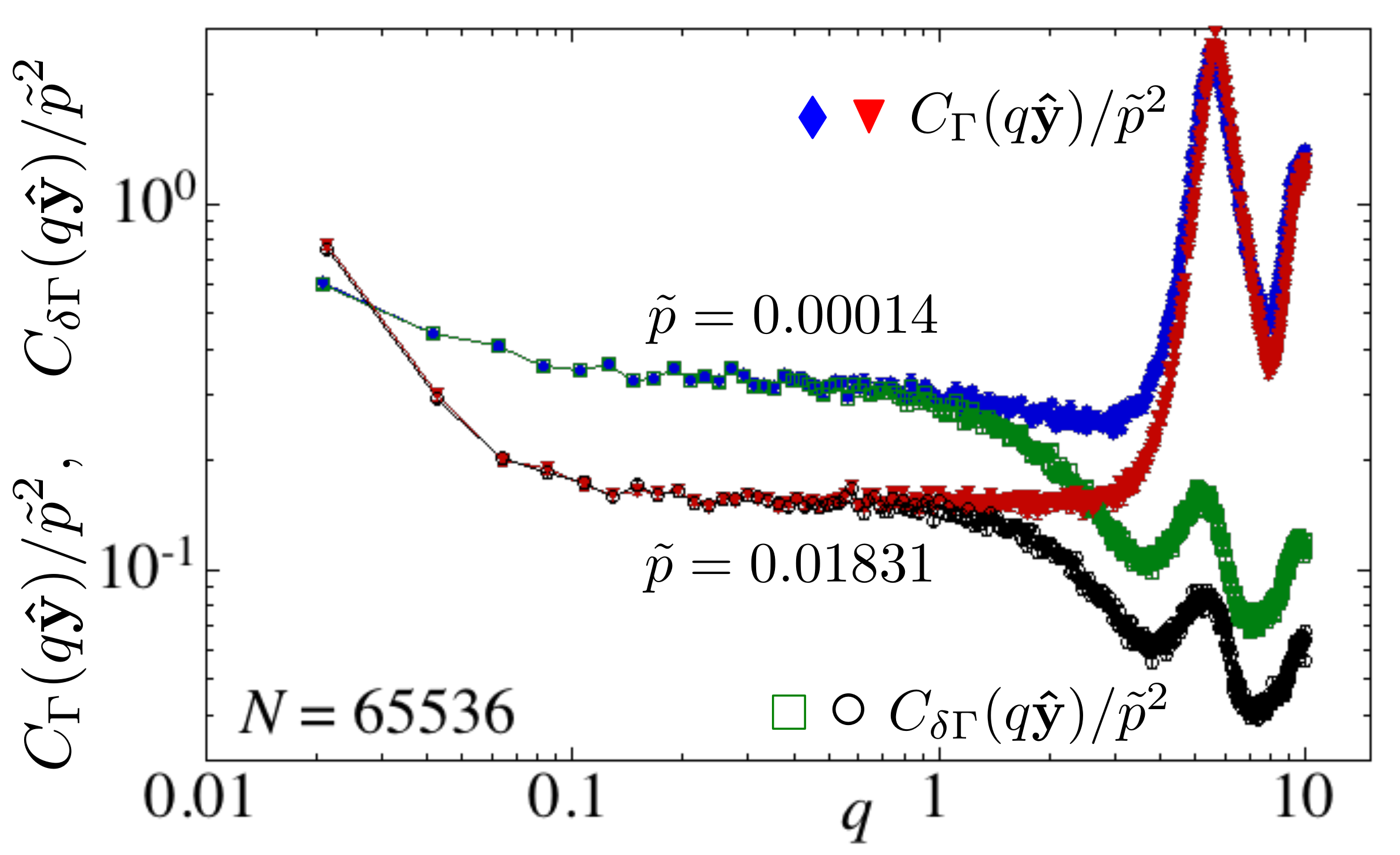}
\caption{(Color online) Comparison of $C_\Gamma(q\mathbf{\hat y})/\tilde p^2$ and $C_{\delta\Gamma}(q\mathbf{\hat y})/\tilde p^2$, vs $q$, for our smallest and largest stress per particle, $\tilde p = 0.00014$ and $0.01831$, respectively.  At both values of $\tilde p$, the two correlations become exactly equal at small $q\lesssim 1$.
}
\label{G-dG}
\end{figure}

Comparing Figs.~\ref{Gq-1200} and \ref{dGq-1200}, or Figs.~\ref{Gqotildep2-vs-q} and \ref{dGqotildep2-vs-q}, we see that the correlations $C_\Gamma(q\mathbf{\hat y})$ and $C_{\delta\Gamma}(q\mathbf{\hat y})$ appear qualitatively the same at small $q$.  In fact, these two correlations are exactly equal at small $q$, as we demonstrate in Fig.~\ref{G-dG}.  From Eq.~(\ref{edecomposition}) we can define the Fourier transforms of the diagonal elements of the stress tensor as,
\begin{equation}
\Sigma_{xx\mathbf{q}}=\Gamma_\mathbf{q}  +\delta\Gamma_\mathbf{q},\quad
\Sigma_{yy\mathbf{q}}=\Gamma_\mathbf{q}  -\delta\Gamma_\mathbf{q}.
\end{equation}
From this we have,
\begin{equation}
\mathrm{covar}(\Sigma_{xx\mathbf{q}}, \Sigma_{yy\mathbf{q}})=\mathrm{var}(\Gamma_\mathbf{q})-\mathrm{var}(\delta\Gamma_\mathbf{q}).
\end{equation}
From the definitions of Eq.~(\ref{eCq}), and the results of Fig.~\ref{G-dG}, we see that $\mathrm{var}(\Gamma_\mathbf{q})=\mathrm{var}(\delta\Gamma_\mathbf{q})$ for $\mathbf{q}=q\mathbf{\hat y}$, and hence $\mathrm{covar}(\Sigma_{xx\mathbf{q}}, \Sigma_{yy\mathbf{q}})=0$.  Note, since $C_\Gamma(\mathbf{q})=\mathrm{var}(\Gamma_\mathbf{q})/V$ is rotationally invariant, and hence independent of the direction of $\mathbf{q}$, while $C_{\delta\Gamma}(\mathbf{q})=\mathrm{var}(\delta\Gamma_\mathbf{q})/V$ depends on the direction of $\mathbf{q}$, this vanishing of $\mathrm{covar}(\Sigma_{xx\mathbf{q}},\Sigma_{yy\mathbf{q}})$ occurs only for the  values of $\mathbf{q}$ that are aligned with the coordinate directions used to define the components of the stress tensor in Eq.~(\ref{edecomposition}), i.e., the $\pm\mathbf{\hat x}$ and $\pm\mathbf{\hat y}$ directions.  For $\mathbf{q}$ in these special directions, the results of Fig.~\ref{G-dG} show that the fluctuations of the diagonal stress elements 
$\Sigma_{xx\mathbf{q}}$ and $\Sigma_{yy\mathbf{q}}$ are statistically independent.
Lema{\^i}tre has recently \cite{Lemaitre2} given theoretical arguments supporting this result based on considerations derived from force balance.

Finally we consider the correlations $C_{\Sigma_{xy}}(q\mathbf{\hat y})$ and $C_{\delta\Gamma}(q\mathbf{\hat e}_{x+y})$.  If fluctuations are isotropic, then according to Eq.~(\ref{rotate2}) these correlations should be equal.  According to the prediction of HC given by Eq.~(\ref{sigHC}), $C_{\Sigma_{xy}}(\mathbf{q})\sim q_x^2q_y^2/q^4$, and so $C_{\Sigma_{xy}}(q\mathbf{\hat y})$  should vanish at any $q$ (since $q_x=0$).  However we find that these correlations, in contrast to the other correlations discussed above, are much more sensitive to the numerical accuracy to which our state is a true energy minimum  obeying exact force balance on each particle.  For our constant stress ensemble of Sec.~\ref{sprotocol} we find we are not able to achieve sufficient accuracy in our energy minimization to accurately compute these correlations at the smallest values of $q$.  However in a fixed volume ensemble we find that we are able to achieve sufficient accuracy at the higher pressures, and we find from these results that 
$C_{\Sigma_{xy}}(q\mathbf{\hat y})=C_{\delta\Gamma}(q\mathbf{\hat e}_{x+y})\sim q^4$.  
Thus we find that the HC prediction, that this correlation should vanish at any $q$, does not hold in general, but rather this correlation only vanishes in the $q\to 0$ limit.  Details of this calculation are discussed in Appendix A.

\subsection{Soft-core particles: Fluctuations in real space}
\label{ssoftR}

\begin{figure}
\includegraphics[width=3.4in]{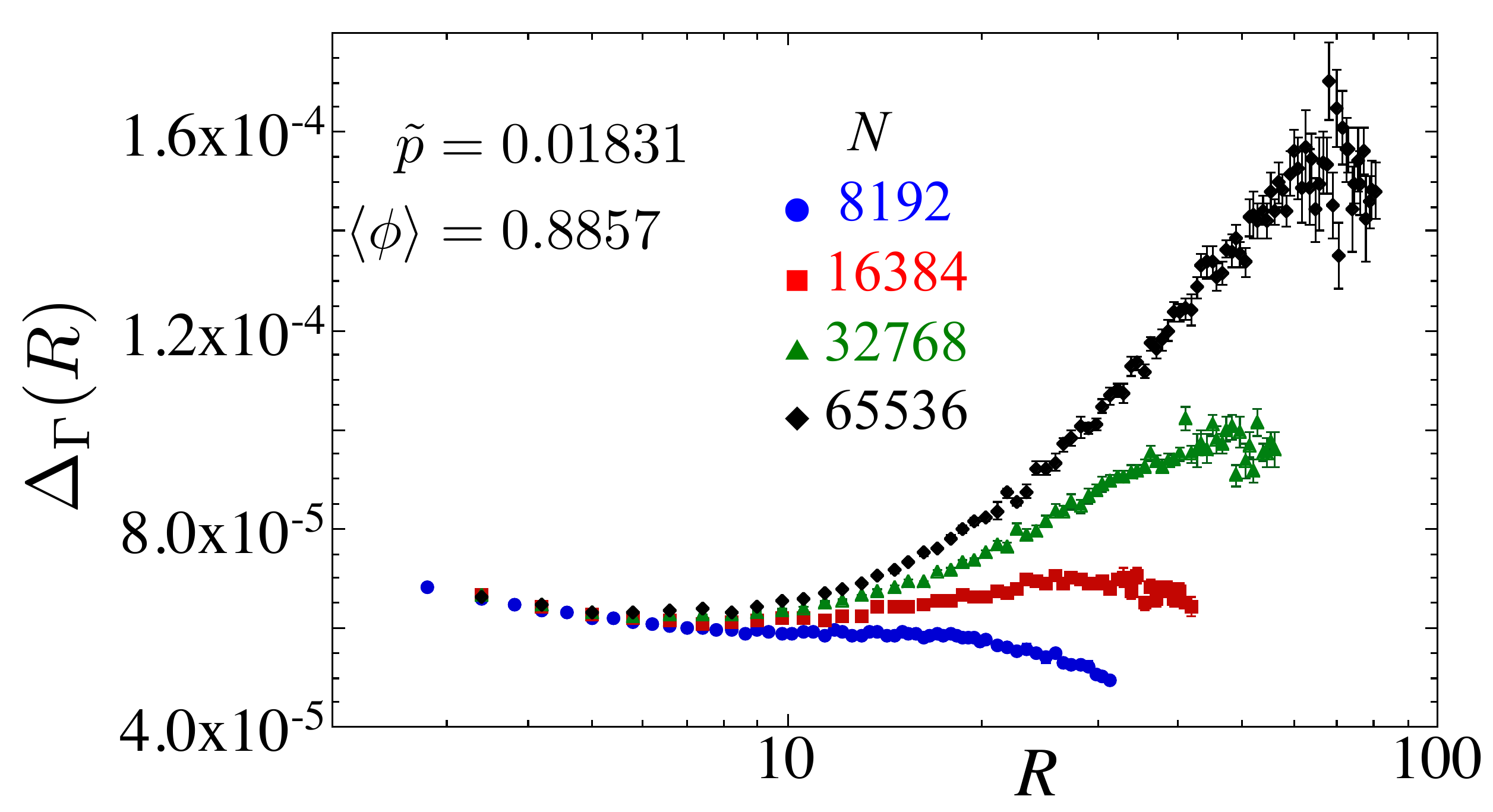}
\caption{(Color online)  Fluctuation of the isotropic part of the stress averaged over a circular window of radius $R$ (see Eq.~(\ref{eDR})), $\Delta_\Gamma(R)$ vs $R$, for systems with different number of particles $N$ at a total stress per particle $\tilde p=0.01831$.
}
\label{varG-1200}
\end{figure}

Here we consider fluctuations of the stress in real space.  We consider first the fluctuations of the isotropic part of the stress $\Gamma$, as measured by the quantity $\Delta_\Gamma(R)$ of Eq.~(\ref{eDR}).
In Fig.~\ref{varG-1200} we plot $\Delta_\Gamma(R)$ vs the window radius $R$ for our largest stress per particle $\tilde  p=\Gamma/N=0.01831$, for system sizes with $N=8192$ to $65536$ particles.  At small $R$, the results for different system sizes agree, and they appear to be approaching a constant value at intermediate lengths $R\sim 8$, consistent with the earlier results of Ref.~\cite{WuTeitel} and as expected if stress fluctuations are self-averaging.  However as $R$ increases further, $\Delta_\Gamma(R)$ starts to increase;  this increase becomes larger as the size of the system $N$ becomes larger.  The fluctuations $\Delta_\Gamma(R)$ reach a maximum and then decrease when the area of the circular window becomes roughly 1/3 the total area of the system, an effect that is due to the periodic Lees-Edwards boundary conditions.

\begin{figure}
\includegraphics[width=3.4in]{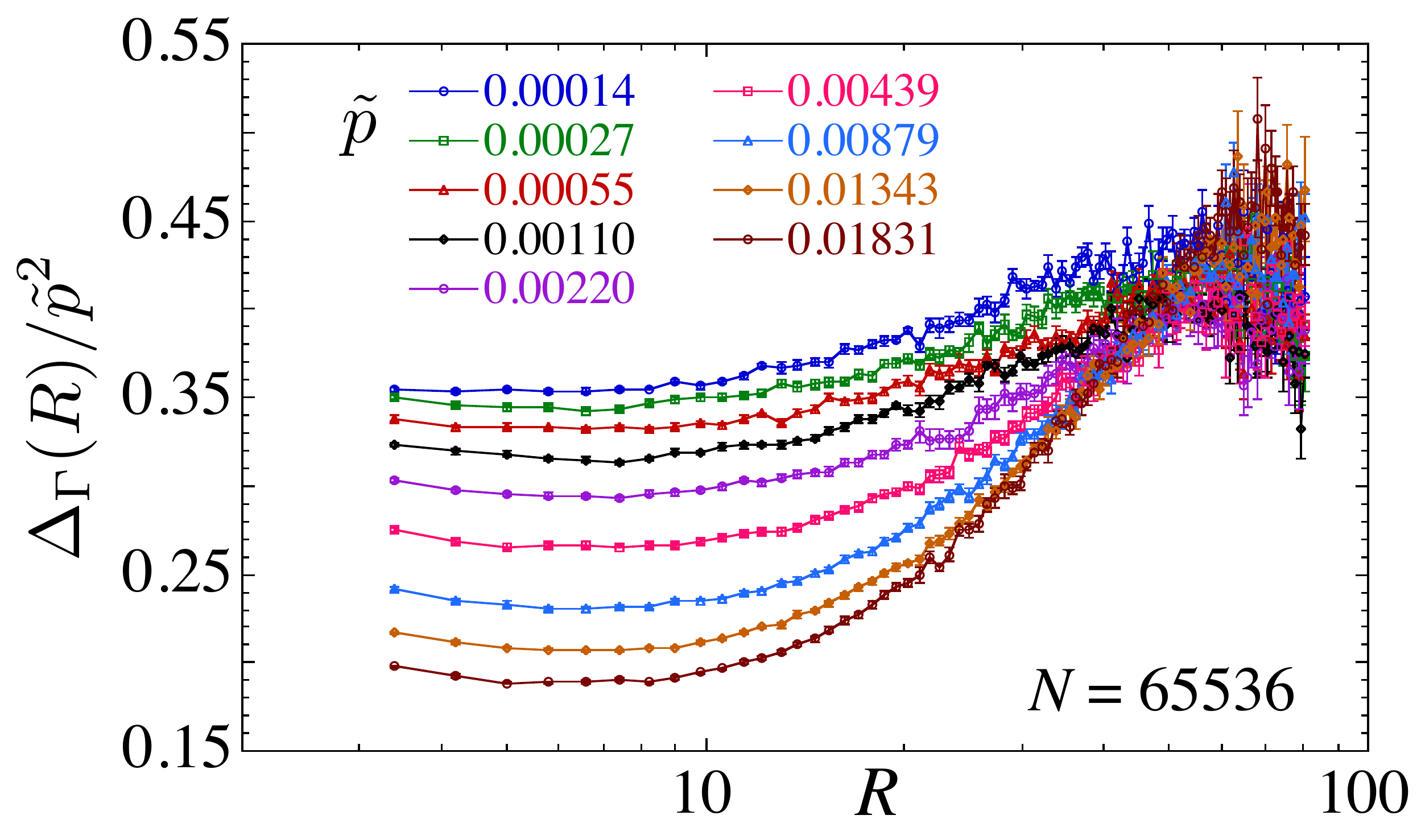}
\caption{(Color online)   Fluctuation of the isotropic part of the stress averaged over a circular window of radius $R$, $\Delta_\Gamma(R)/\tilde p^2$ vs $R$.
Curves are for different values of the total system stress per particle $\tilde p = \Gamma/N$, going from $\tilde p = 0.00014$ on top to $0.01831$ on bottom. System has $N = 65536$ particles.
}
\label{vGRotildep2}
\end{figure}

The marked finite size effect seen for $\Delta_\Gamma(R)$ in Fig.~\ref{varG-1200} should be contrasted with the absence of any finite size effect found for $C_\Gamma(\mathbf{q})$ in Fig.~\ref{Gq-1200}.  This leads one to conclude that the finite size effect in $\Delta_\Gamma(R)$ as $N$ varies must be due to the difference in the allowed set of $\{\mathbf{q}\}$ values that appear in the sum of Eq.~(\ref{eDC}).  Since these allowed $\{\mathbf{q}\}$ are $\mathbf{q}=2\pi[(m_1/L_1)\mathbf{\hat x}+(m_2/L_2-\gamma m_1/L_1)\mathbf{\hat y}]$, $m_1,m_2$ integer, the bigger the system size $N$ (and hence the larger the system length $L$), the smaller are the $|\mathbf{q}|$ that enter the sum in Eq.~(\ref{eDC}); since $C_\Gamma(\mathbf{q})$ is growing at small $|\mathbf{q}|$, the contribution from ever smaller $|\mathbf{q}|$ as $N$ increases, gives rise to the finite size effect seen in Fig.~\ref{varG-1200}.
In Fig.~\ref{vGRotildep2} we plot $\Delta_\Gamma(R)/\tilde p^2$ vs $R$ for our largest system with $N=65536$ particles, showing results for a range of total stress per particle $\tilde p$.  We see that the growth in the large $R$ fluctuations gets more pronounced as $\tilde p$ increases.

\begin{figure}[h!]
\includegraphics[width=3.4in]{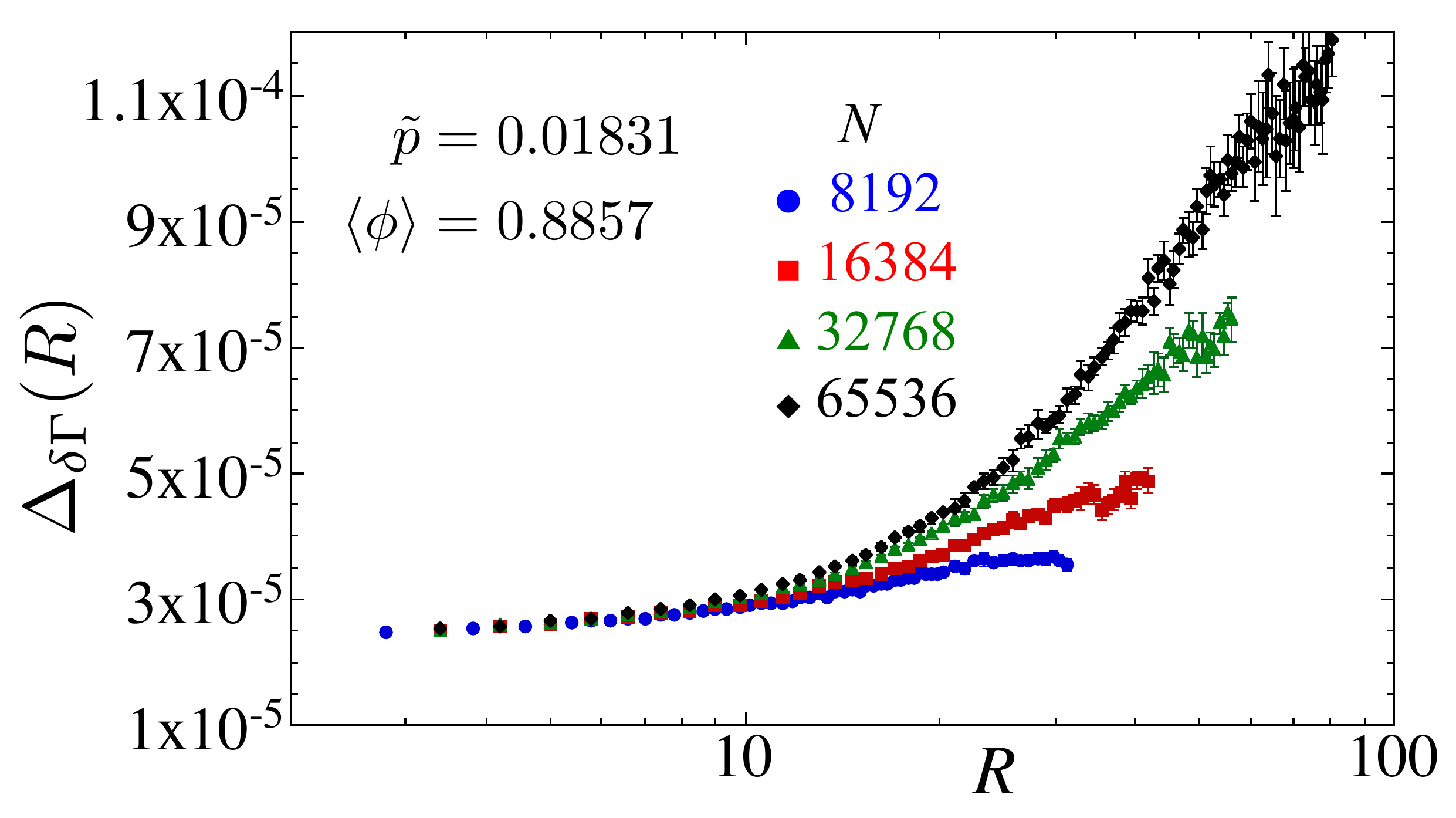}
\caption{(Color online)  Fluctuation of the anisotropic part of the stress $\delta\Gamma$ averaged over a circular window of radius $R$ (see Eq.~(\ref{eDR})), $\Delta_{\delta\Gamma}(R)$ vs $R$, for systems with different number of particles $N$ at a total stress per particle $\tilde p=0.01831$.
}
\label{vardG-1200}
\end{figure}

\begin{figure}[h!]
\includegraphics[width=3.4in]{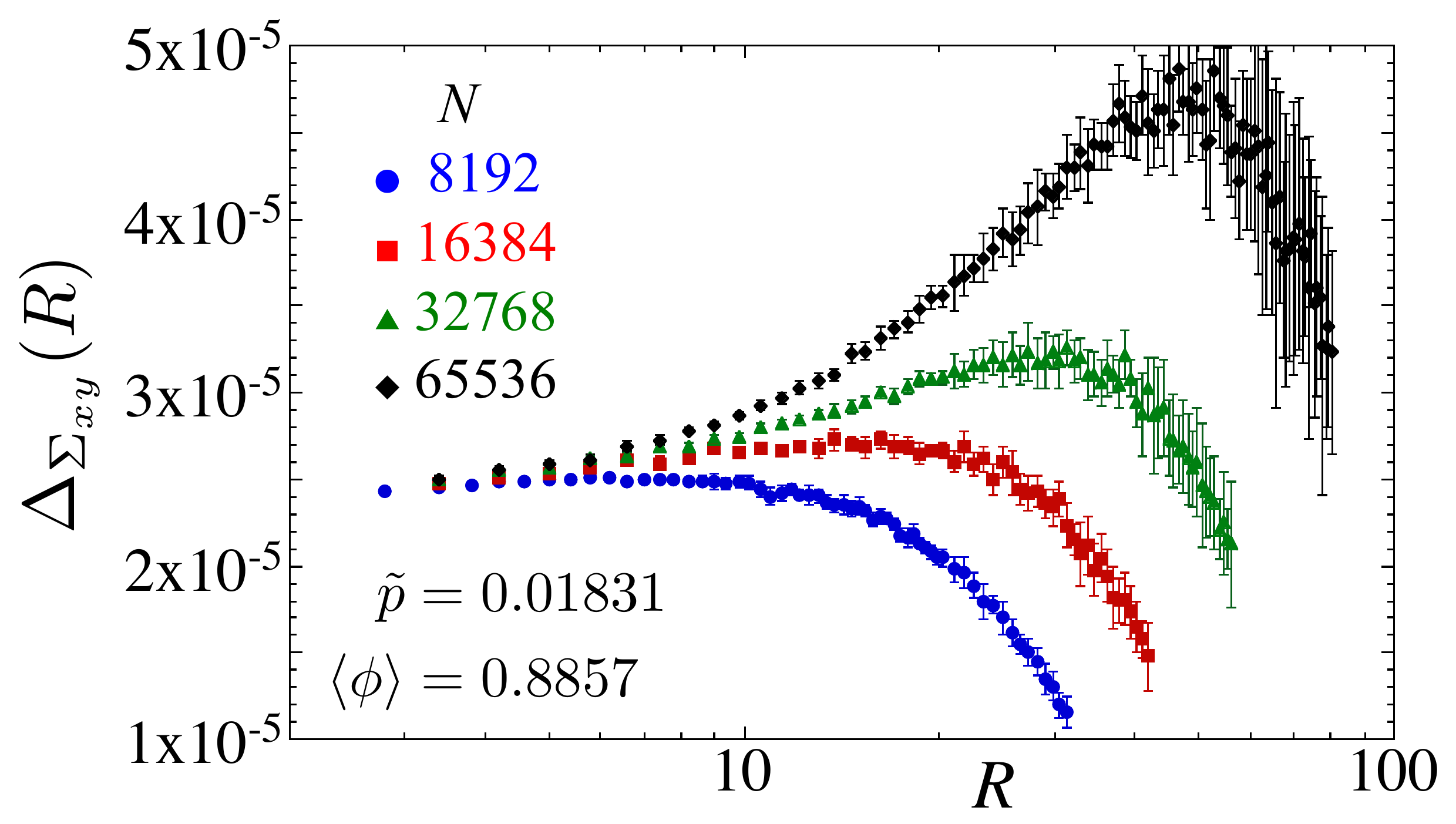}
\caption{(Color online)  Fluctuation of the anisotropic part of the stress $\Sigma_{xy}$ averaged over a circular window of radius $R$ (see Eq.~(\ref{eDR})), $\Delta_{\Sigma_{xy}}(R)$ vs $R$, for  systems with different number of particles $N$ at a total stress per particle $\tilde p=0.01831$.
}
\label{varGxy-1200}
\end{figure}

In Figs.~\ref{vardG-1200} and \ref{varGxy-1200} we plot the fluctuation of the anisotropic parts of the stress tensor, as measured by $\Delta_{\delta\Gamma}(R)$ and $\Delta_{\Sigma_{xy}}(R)$ vs $R$, for several different system sizes $N$ at $\tilde p=0.01831$.  Again we see that at {\em small} $R\lesssim 8$ there is little dependence on the system size $N$, the fluctuations appear roughly constant in $R$, and the fluctuations of $\delta\Gamma$ and $\Sigma_{xy}$ are equal, as would be expected if the fluctuations are isotropic and self-averaging.  However as $R$ increases, we see a significant dependence on the system size, and the fluctuations of $\Sigma_{xy}$ become significantly smaller than those of $\delta\Gamma$.
Our results here look qualitatively similar to those for the deviatoric stress shown in Ref.~\cite{Karimi}.

\begin{figure}[h!]
\includegraphics[width=3.4in]{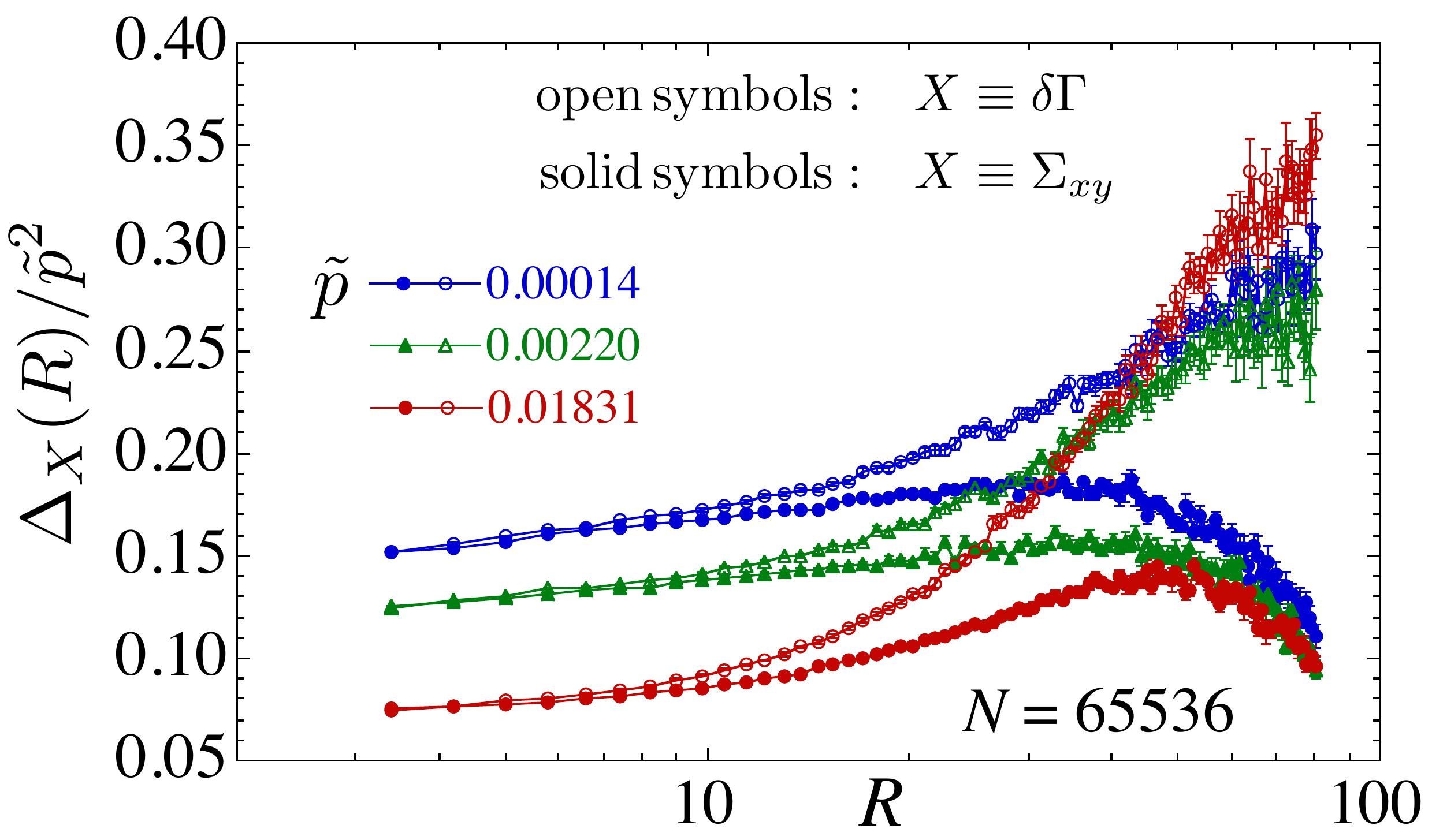}
\caption{(Color online)  Fluctuation of the anisotropic parts of the system stress averaged over a circular window of radius $R$, $\Delta_{\delta\Gamma}(R)/\tilde p^2$ and $\Delta_{\Sigma_{xy}}(R)/\tilde p^2$ vs $R$, for three different values of the total system stress per particle $\tilde p$.
System has $N = 65536$ particles.
}
\label{vdGR-and-vGxyR-otildep2}
\end{figure}

To illustrate the breaking of rotational isotropy of spatial fluctuations at large $R$, in Fig.~\ref{vdGR-and-vGxyR-otildep2} we plot both $\Delta_{\delta\Gamma}(R)/\tilde p^2$ and $\Delta_{\Sigma_{xy}}(R)/\tilde p^2$ vs $R$ at our smallest, largest, and an intermediate value of $\tilde p$, for our largest system with $N=65536$ particles.  We see clearly that the rotational isotropy at small lengths scales $R$, characterized by $\Delta_{\delta\Gamma}(R)=\Delta_{\Sigma_{xy}}(R)$, breaks down as $R$ increases.  
This break down of rotational isotropy at large $R$ presumably occurs when the circular window of averaging becomes a sizable fraction of the total system box, since the system box itself (see Fig.~\ref{box}) is not rotationally isotropic.

\begin{figure}[h!]
\includegraphics[width=3.4in]{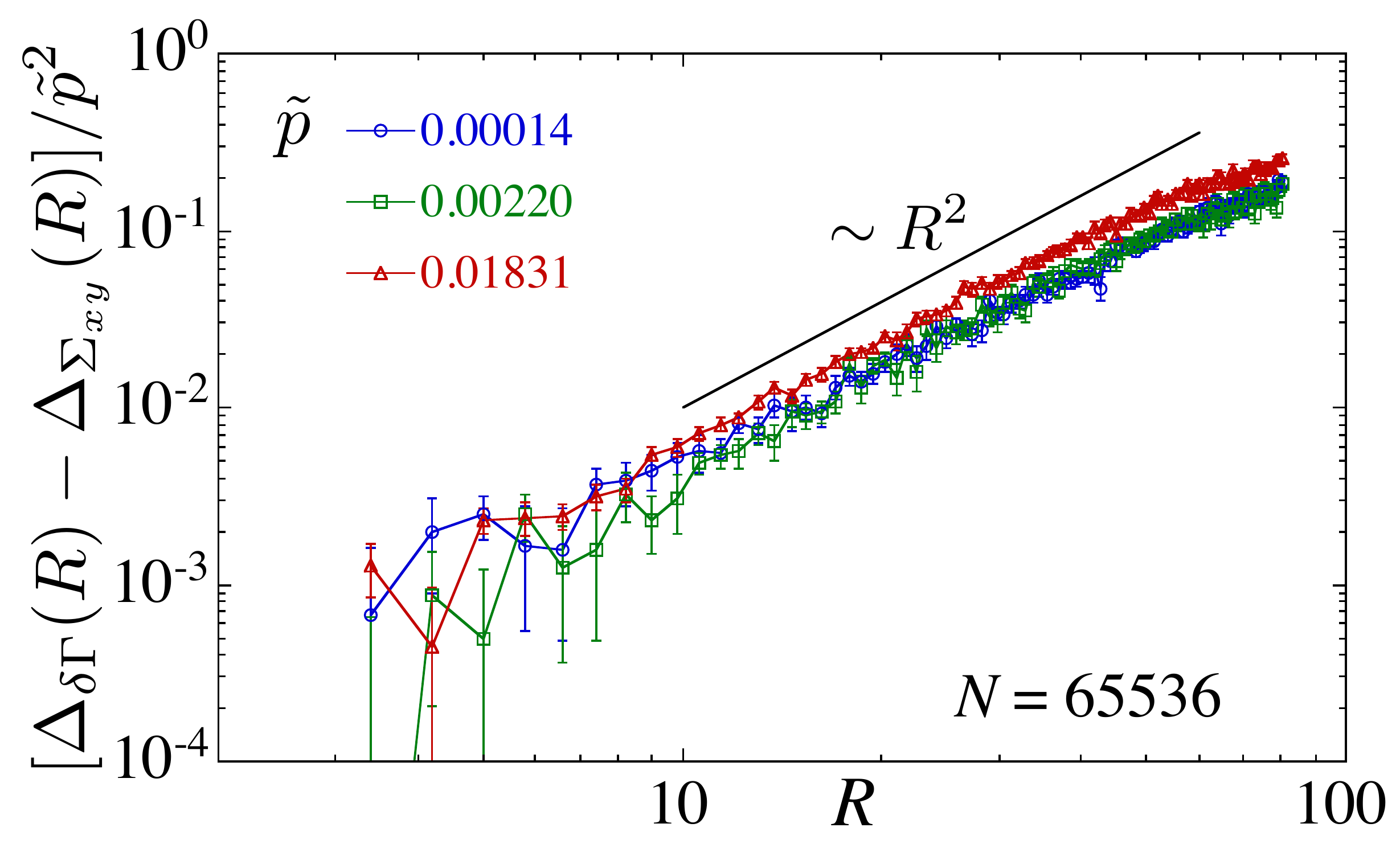}
\caption{(Color online) Difference $[\Delta_{\delta\Gamma}(R)-\Delta_{\Sigma_{xy}}(R)]/\tilde p^2$ vs $R$ for three different values of the total system stress per particle $\tilde p$. System has $N = 65536$ particles.  Solid line has slope $2$.
}
\label{vdGR-vGxyR-otildep2}
\end{figure}

One might think that it could be possible to define a length scale characterizing this break down in the isotropy of spatially averaged fluctuations.  To check this, in Fig.~\ref{vdGR-vGxyR-otildep2} we plot the difference $[\Delta_{\delta\Gamma}(R)-\Delta_{\Sigma_{xy}}(R)]/\tilde p^2$ vs $R$.  We see that this difference scales algebraically with $R$ (roughly $\sim R^2$), rather than defining any obvious length scale.  
To conclude, our results in this section show explicitly that the spatial fluctuation measures $\Delta_X(R)$ do not become constant as $R$ increases, but rather increase with increasing $R$, again demonstrating that the stress fluctuations are not self-averaging.

\subsection{Soft-core particles: Testing protocol dependence}
\label{sprotocol2}

It is known that, when constructing jammed packings of frictionless disks by compression or quenching, the location of the critical packing fraction of the jamming transition $\phi_J$, below which mechanically stable packings no longer exist and the stress vanishes, may be sensitive to the details of the particular protocol used to construct the mechanically stable packings \cite{Chaudhuri,VOT}.  Although other quantities, such as the exponents that describe the vanishing of pressure and elastic moduli as $\phi\to\phi_J$ from above, seem to be independent of protocol \cite{Chaudhuri}, one may still question whether the anomalous large length scale stress fluctuations we find in the present work might not be some artifact of our particular protocol.

In particular, when deriving packings by quenching (rapid energy minimization) at fixed volume, the ensemble of mechanically stable configurations that one finds can depend on the ensemble of initial configurations that one quenches from \cite{VOT}.  Hence one may wonder if the results reported in the previous sections might not depend on the value of $\phi_\mathrm{init}=0.84$, which we took as the density of our initial random configurations, see  Eq.~(\ref{ephinit}); the value 0.84 is just slightly below the $\phi_J=0.8416$ for our protocol.  Such dense initial random configurations contain many particles with large overlaps and one may wonder if the large density fluctuations of these initial configurations somehow get frozen in during the quenching process.  

\begin{figure}
\includegraphics[width=3.4in]{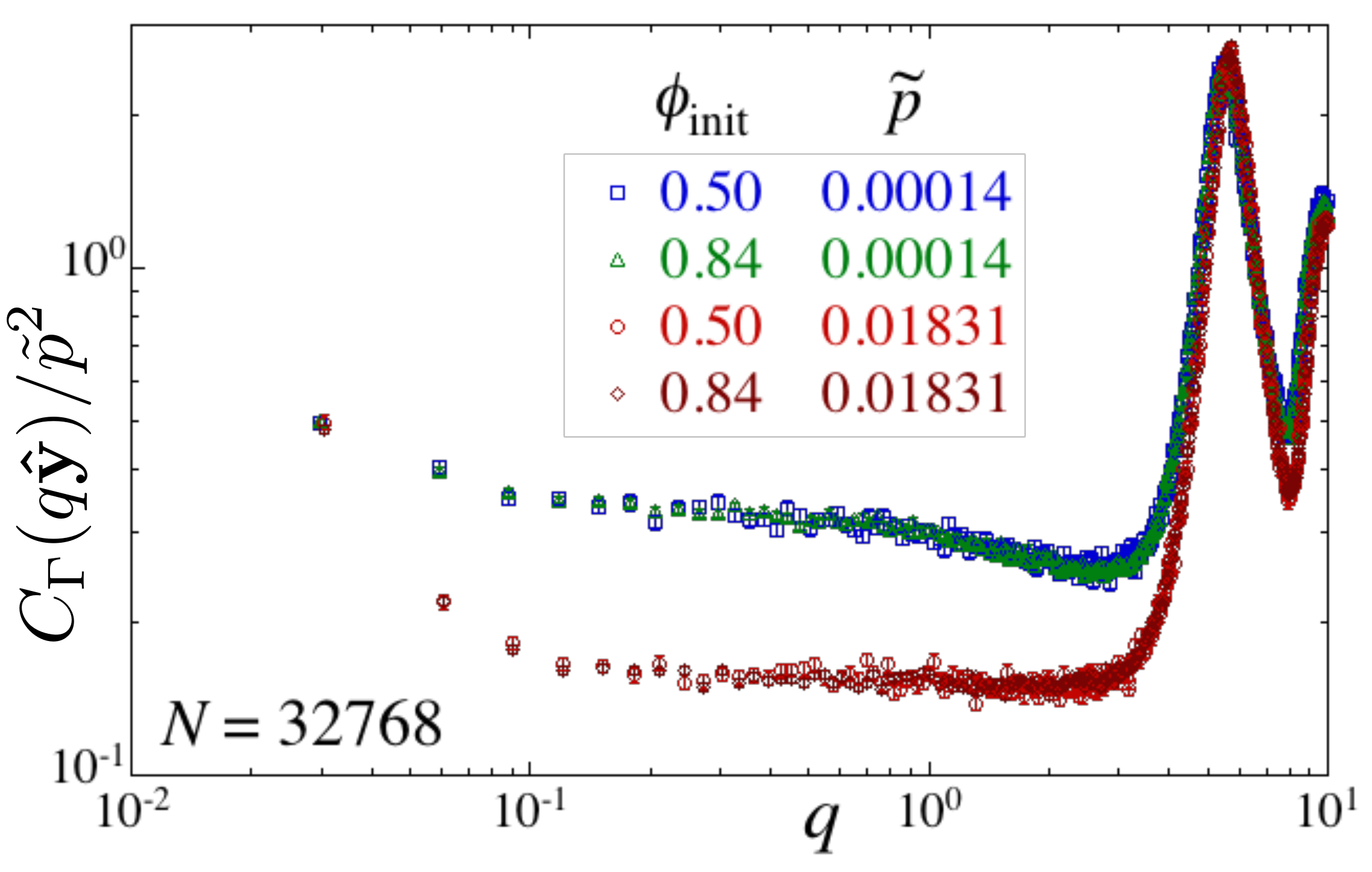}
\caption{(Color online) Fluctuation of the isotropic part of the stress $C_\Gamma(q\mathbf{\hat y})/\tilde p^2$ vs $q$, comparing results obtained when quenching from initial random configurations with $\phi_\mathrm{init}=0.84$ with those obtained from $\phi_\mathrm{init}=0.50$.  Results are shown for final configurations at two different values of the stress per particle $\tilde p = \Gamma/N = 0.00014$ and 0.01831.  The system has $N=32768$ particles and results are averaged over roughly 1000 independent initial configurations.
}
\label{Gq-phi_init}
\end{figure}

To test this we have also constructed mechanically stable packings by starting from initial random configurations with the much smaller packing fraction $\phi_\mathrm{init}=0.50$.  In Fig.~\ref{Gq-phi_init} we show results for the correlation of the isotropic part of the stress $C_\Gamma(q\mathbf{\hat y})$ vs $q$, comparing results from $\phi_\mathrm{init}=0.84$ with those from $\phi_\mathrm{init}=0.50$, at our smallest and our largest values of the stress per particle $\tilde p$.  We find essentially no dependence at all on the value of $\phi_\mathrm{init}$.  A similar agreement is found for the correlations $C_{\delta\Gamma}(q\mathbf{\hat y})$ and $C_{\Sigma_{xy}}(q\mathbf{\hat y})$.  We also find that the average packing fraction $\langle\phi\rangle$ as a function of $\tilde p$ shows no dependence on $\phi_\mathrm{init}$.

In retrospect, the independence of our results on $\phi_\mathrm{init}$ is not surprising.  Recall that our protocol of Sec.~\ref{sprotocol} varies both particle positions and box size and shape, so as to minimize $\tilde U$ of Eq.~(\ref{eUtilde}) to a target value of $\Gamma$.   When we start with an initial large $\phi_\mathrm{init}=0.84$, we find that the first few steps of our minimization algorithm give a rapid increase of the box size to reach a relatively low packing fraction, as the initially overlapping particles push away from each other; once the particles have spread out to reduce the overlaps to negligible amounts, only then does the algorithm start to compress the box to achieve the target value of total stress $\Gamma$ (this occurs automatically with our conjugate gradient minimization algorithm; it is not something put in by hand).

\begin{figure}[h!]
\includegraphics[width=3.4in]{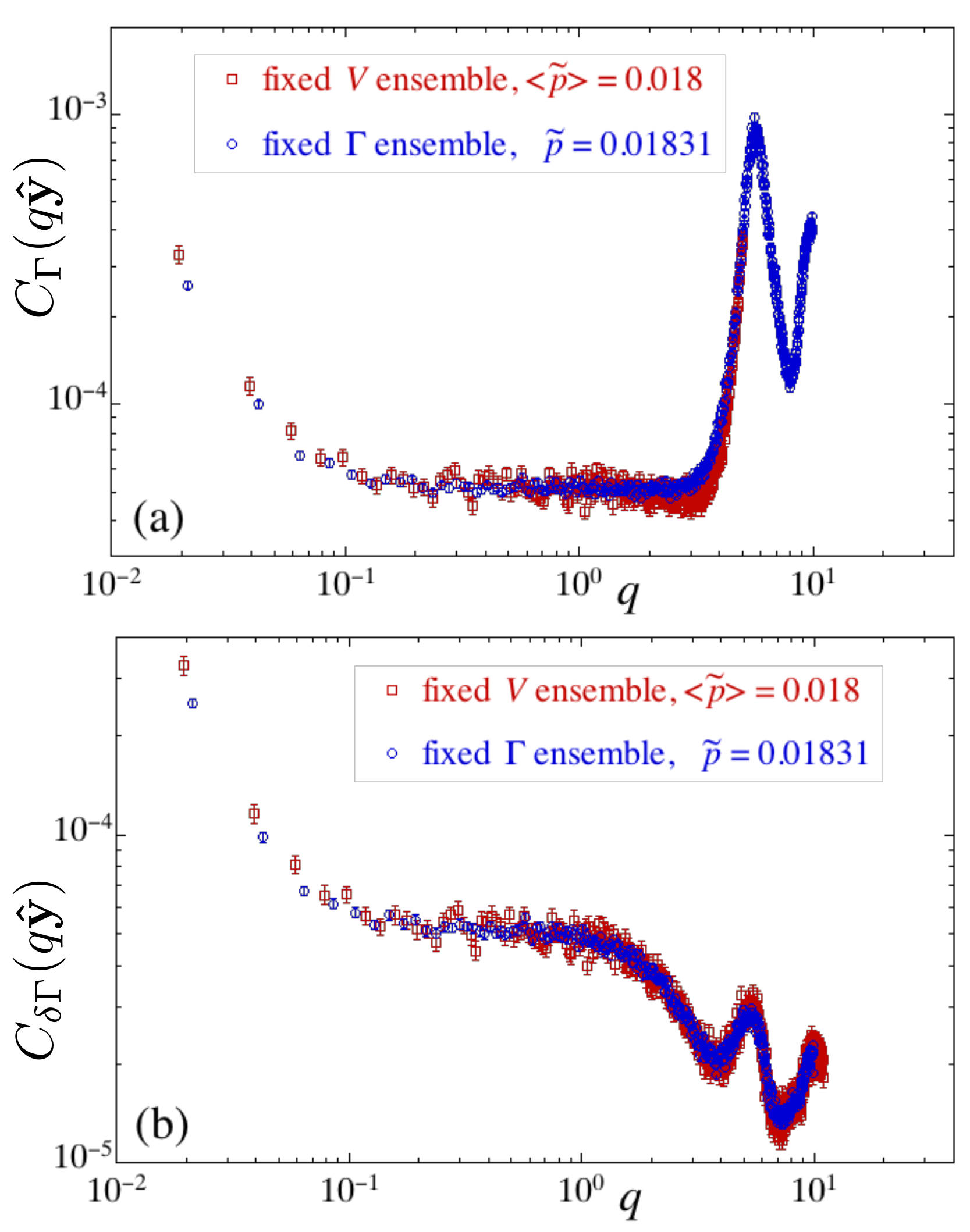}
\caption{(Color online) Fluctuation of the (a) isotropic part of the stress $C_\Gamma(q\mathbf{\hat y})$ 
and (b) anisotropic part of the stress $C_{\delta\Gamma}(q\mathbf{\hat y})$
vs $q$.  Circles are for an ensemble of $N=65536$ particles at fixed stress, with $\tilde p=\Gamma/N=0.01831$ and $\langle\phi\rangle=0.8857$.  Squares are for an ensemble of $N= 77523$ particles at fixed square volume of side length $L=320$, with $\langle \tilde p\rangle=0.018$ and $\phi=0.88$.
}
\label{compare}
\end{figure}

To further demonstrate that the increasing stress fluctuations which we find as $q\to 0$ are not somehow an artifact of our particular fixed stress protocol, we have also constructed mechanically stable packings by quenching from random initial configurations at fixed volume \cite{Karimi3}.  In this case we start with random particle configurations in a square box of length $L$, and then minimize the total elastic energy $U$ to find mechanically stable final configurations, keeping box size and shape fixed.  The packing fraction $\phi$ remains constant throughout this process.  The final configurations produced by this method may contain some residual total shear stress.  However this  residual shear stress, relative to the isotropic part $\Gamma$, scales as the inverse square root of the system size, and so for our very large systems with $N=77523$ particles it is completely negligible.
In Fig.~\ref{compare} we plot the resulting $C_\Gamma(q\mathbf{\hat y})$ 
and $C_{\delta\Gamma}(q\mathbf{\hat y})$
vs $q$ for a system of fixed length $L=320$ at packing fraction $\phi=0.88$.  Our system has an average stress per particle of   $\langle \tilde p\rangle =\Gamma/N=0.018$.  Our results are averaged over 256 independent configurations.  
In the same figure we show our results from Figs.~\ref{Gq-1200} and \ref{dGq-1200} for the fixed stress ensemble with $N=65536$ particles, $\tilde p=0.01831$ and $\langle \phi\rangle=0.8857$.  We see quite consistent agreement, given the small difference in the values of $\tilde p$.  Our results thus show that the anomalous small $q$ stress fluctuations found for these two correlations are robust, rather than an artifact of the particular protocol used to construct our mechanically stable packings.

\subsection{Inherent states of a Lennard-Jones liquid: Correlations in $\mathbf{q}$-space}
\label{sLJ}

In addition to the soft-core harmonically repelling disks that are the main focus of the present work, we have found similar anomalous large length scale stress fluctuations in the inherent states of a dense binary Lennard-Jones (LJ) liquid.  We consider a LJ liquid with equal numbers of ``small" and ``big" particles, with effective diameters $d_s$ and $d_b$ respectively, with $d_b/d_s=1.4$.  We will measure lengths in units where $d_s=1$.

\begin{figure}
\includegraphics[width=3.4in]{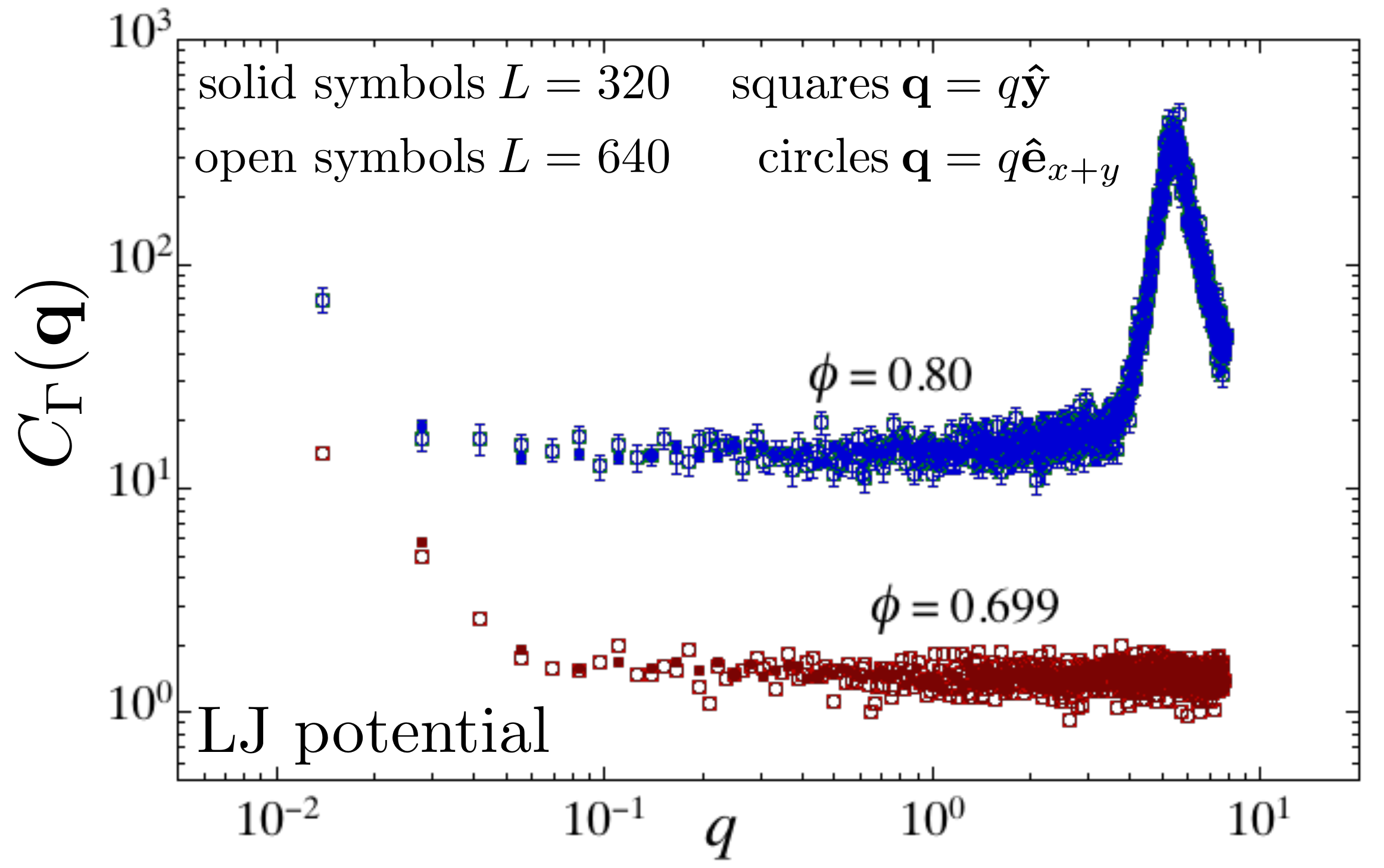}
\caption{(Color online) Fluctuation of the isotropic part of the stress $C_\Gamma(\mathbf{q})$ vs $q$ for $\mathbf{q}=q\mathbf{\hat y}$ (squares) and for $\mathbf{q}=q\mathbf{\hat e}_{x+y}$ (circles), for a binary Lennard-Jones system with effective packing fractions $\phi=0.699$ and $0.80$, in a fixed volume ensemble.  Solid symbols are for a system of length $L=320$, open symbols are for $L=640$.  Lengths are measured in units where small particles have diameter $d_s=1$, while big particles have $d_b=1.4$.  
}
\label{LJ}
\end{figure}

Particle $i$ interacts with particle $j$ according to the usual LJ potential,
\begin{equation}
{\cal V}_{ij}(r)=4\varepsilon\left[\left(\frac{d_{ij}}{r}\right)^{12}-\left(\frac{d_{ij}}{r}\right)^6\right],
\end{equation}
where $d_{ij}=  (d_i+d_j)/2$, and $r$ is the  center-to-center distance between the particles \cite{LJnote}. We take the unit of energy such that $\varepsilon=1$. Since ${\cal V}_{ij}(r)=0$ when $r=d_{ij}$, we can view the LJ potential as a soft-core repulsion for particles with diameters $d_s$ and $d_b$, together with a short ranged attractive tail.  We can thus define the effective packing fraction for $N$ particles in a fixed square box of length $L$ to be,
\begin{equation}
\phi = \frac{N}{L^2}\frac{\pi}{2}\left[\left(\frac{d_s}{2}\right)^2+\left(\frac{d_b}{2}\right)^2\right].
\end{equation}
Starting from random particle positions at a fixed $\phi$, we quench at constant volume to a local energy minimum of the LJ potential, to find the ``inherent states" of the LJ system.  

We consider here two different packing fractions, $\phi=0.699$ with $\langle\tilde p\rangle\approx0.05$ and $\phi=0.80$ with $\langle\tilde p\rangle\approx 8.0$.  
The first case corresponds to an average separation between particles of $s=L/\sqrt{N}=1.29$, while the second case has $s=1.2$.  For comparison, the minimum of the LJ potential between two particles $i$ and $j$ lies at $r_0=2^{1/6}d_{ij}\approx 1.12d_{ij}$.

In Fig.~\ref{LJ} we plot the resulting correlation for the isotropic part of the stress fluctuations, $C_\Gamma(\mathbf{q})$ vs $q$, for $\mathbf{q}$ in both the $\mathbf{\hat y}$ and $\mathbf{\hat e}_{x+y}$ directions, for these two values of $\phi$. We show results for two different system sizes, $L=320$  (averaged over 256 independent configurations) and $L=640$ (averaged over 64 independent configurations).  For $\phi=0.699$ these sizes correspond to $N=61568$ and $246272$ particles respectively; for $\phi=0.80$ we have $N=70476$ and $281902$.
As for the case of harmonically repelling soft-core disks, we find that the stress fluctuations are isotropic and take a dramatic turn upwards as $q$ decreases below a finite $q_0$, and that this effect does not depend on the system size.  Unlike with the harmonic disks, we see a noticeable increase in $q_0$ (and so a {\em decrease} in the length scale $\ell_0\approx 2\pi/q_0$)  as $\langle \tilde p\rangle$ decreases.

\subsection{Testing for scaling}
\label{sscaling}

Our analysis of stress correlations for soft-core interacting disks has demonstrated that there is a length scale $\ell_0$, roughly 60 particle diameters long, beyond which stress fluctuations are anomalously large and lead to a breakdown of stress self-averaging.  It is natural to wonder if this large length $\ell_0$ is in some way related to the diverging length scales associated with the jamming transition.

\begin{figure}
\includegraphics[width=3.2in]{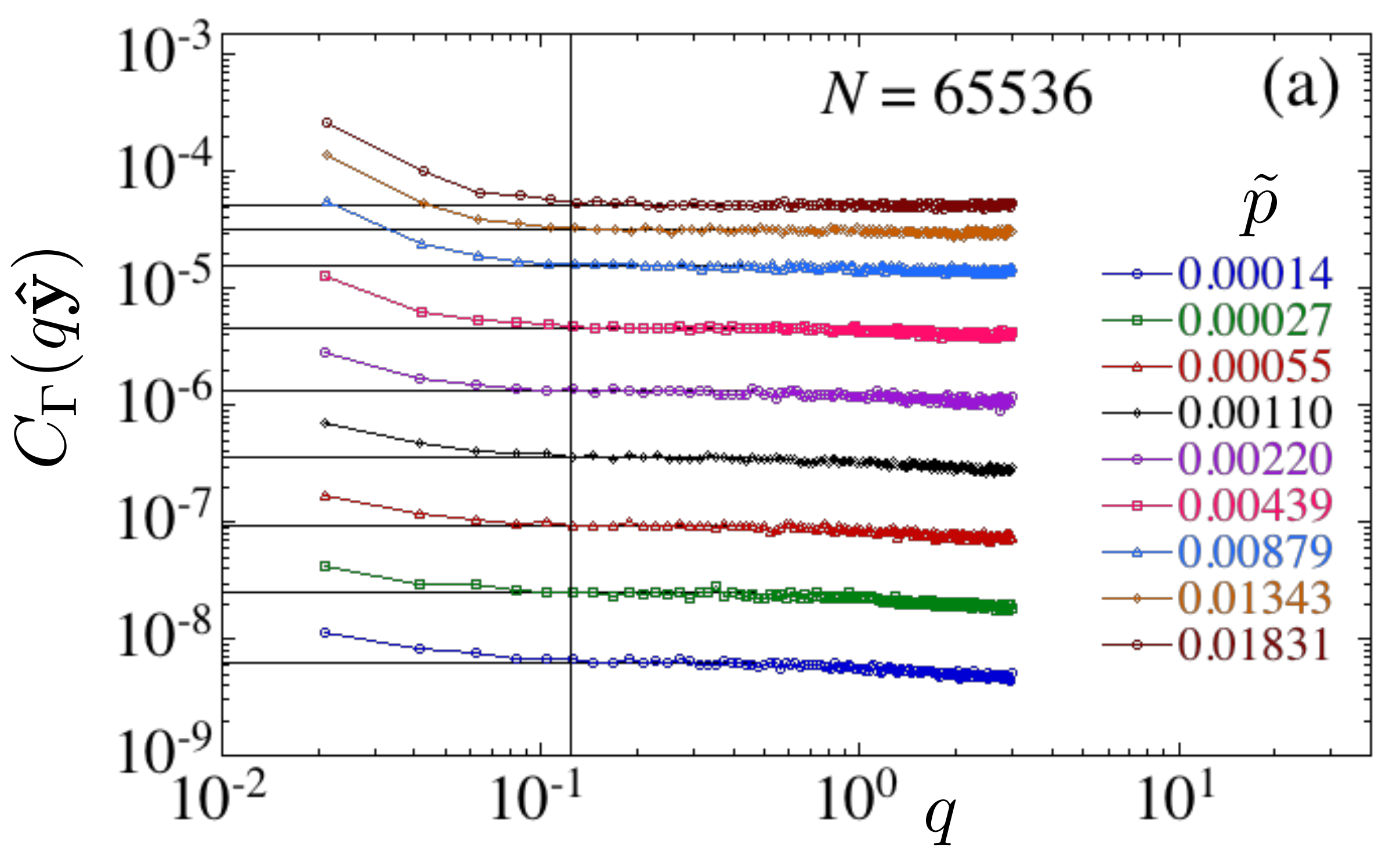}
\includegraphics[width=3.2in]{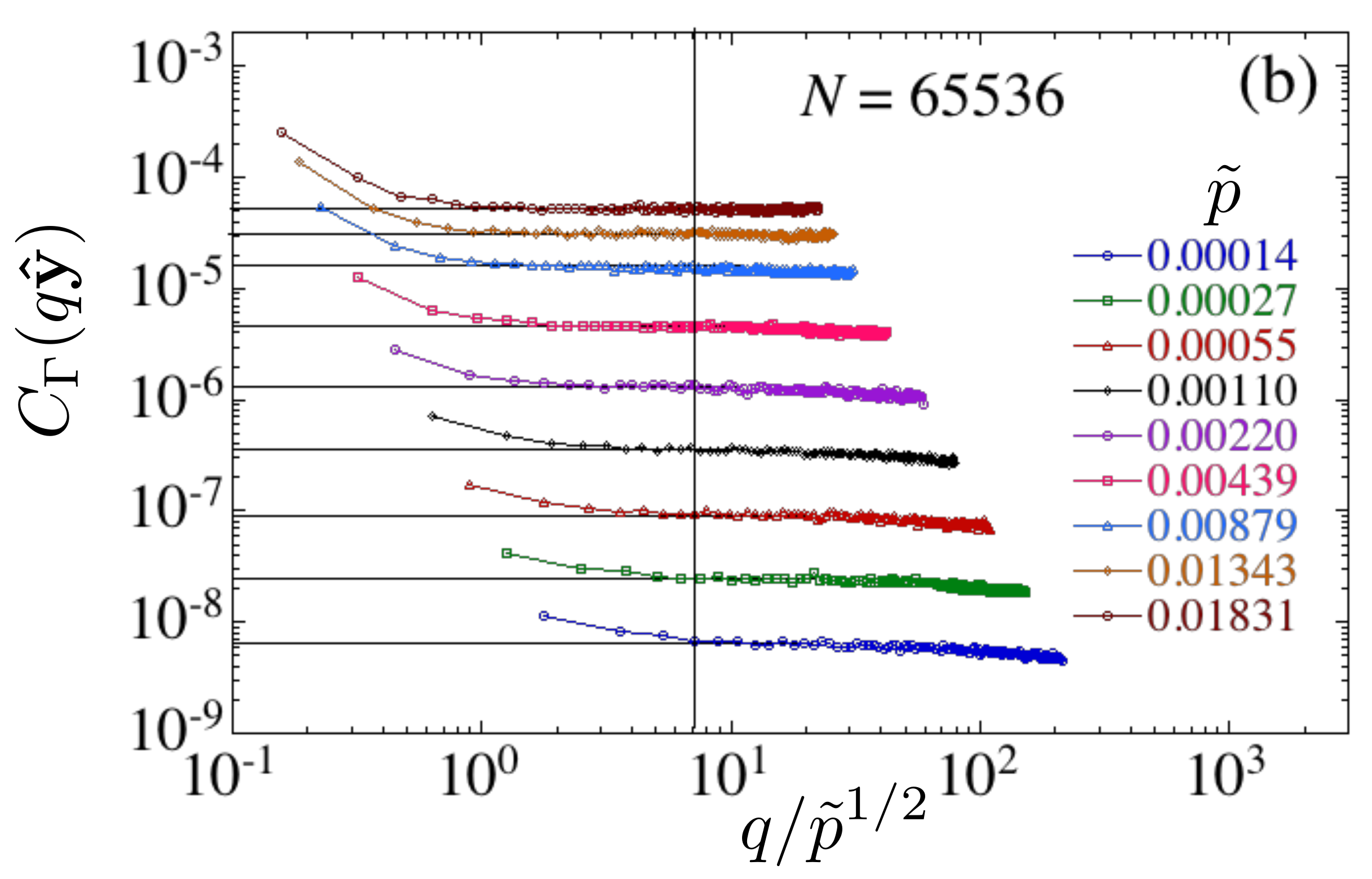}
\includegraphics[width=3.2in]{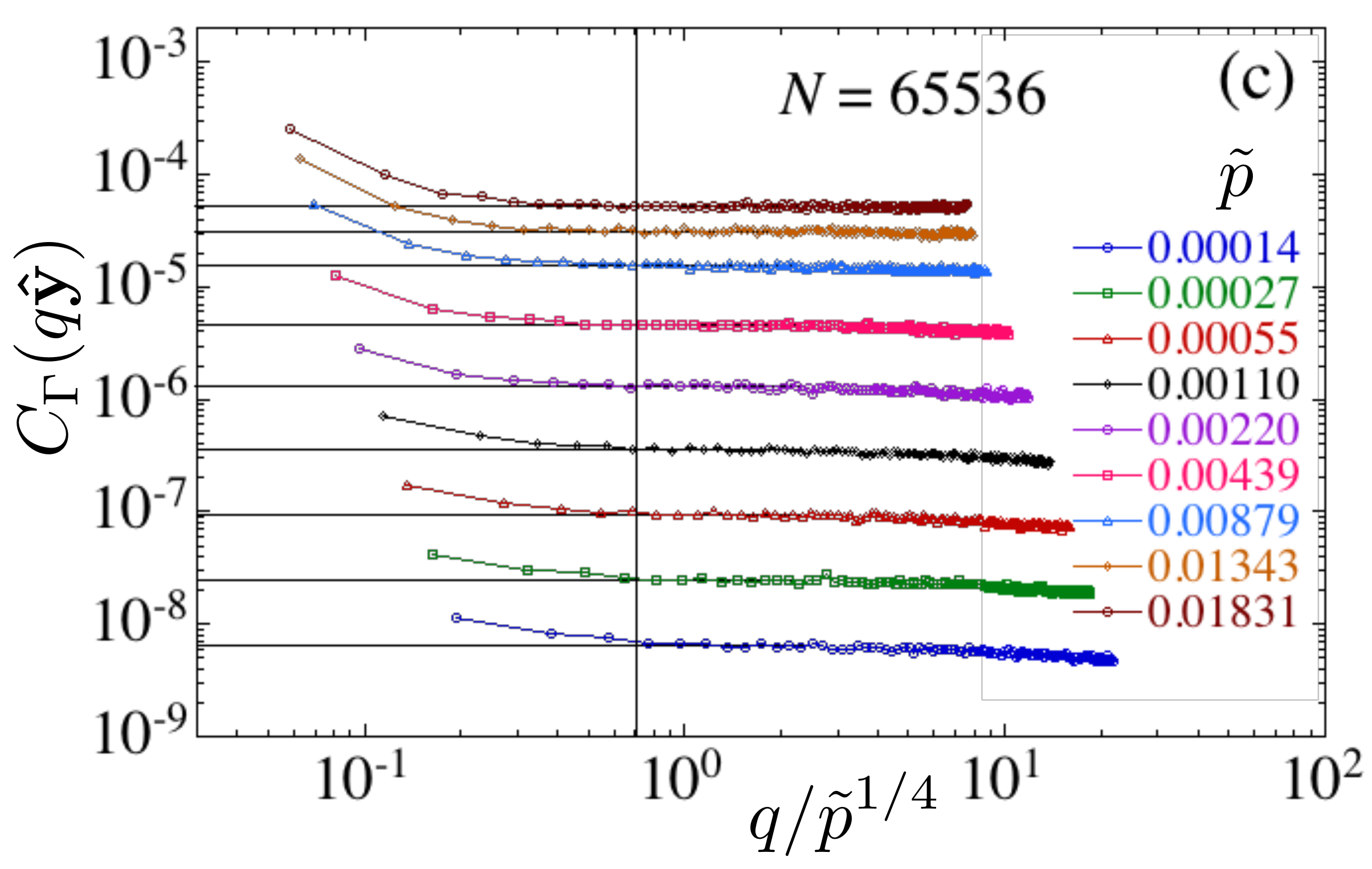}
\caption{(Color online) Plot of $C_\Gamma(q\mathbf{\hat y})=\langle\Gamma_\mathbf{q}\Gamma_{-\mathbf{q}}\rangle/V$ vs (a) $q$, (b) $q/\tilde p^{1/2}\sim q\ell_L$, and (c) $q/\tilde p^{1/4}\sim q\ell_T$, for different values of the total stress per particle $\tilde p=\Gamma/N$ for a system with $N=65536$ particles.  Curves are for $\tilde p = 0.01831$ to $0.00014$ going top to bottom.  The high$-q$ data, where one sees the peak in Fig.~\ref{Gqotildep2-vs-q}, have been truncated since if there is critical scaling it would apply only to the long length scale, and so small$-q$, region of the data. Horizontal solid lines extrapolate through the region where the curves $C_\Gamma(q\mathbf{\hat y})$ are approximately constant; the solid vertical line denotes the approximate point where $C_\Gamma(q\mathbf{\hat y})$ departs from this horizontal line, for the smallest $\tilde p=0.00014$.
}
\label{Gq-vs-qell}
\end{figure}

For our system of soft-core interacting disks, as the stress per particle $\tilde p$ decreases towards zero, the average packing fraction $\phi$ approaches a value $\phi_J$, known as the jamming transition \cite{OHern,Liu+Nagel,vanHecke,Wyart0}.
Exactly at this jamming transition for frictionless spherical particles, the system is isostatic, and the average number of contacts per particle $z$ is $z_c=2d$, with $d$ the dimensionality of the system.  Increasing $\tilde p$ to finite values
above the jamming transition, the average contact number $z$ increases.
Wyart et al. \cite{Wyart} showed how this increase of contacts, $\delta z= z-z_c$, leads to an isostatic length scale  $\ell^*\sim 1/\delta z$, that therefore diverges as the jamming transition is approached from above.  By consideration of the density of soft elastic modes in mechanically stable packings above jamming, Silbert et al. \cite{Silbert} and Wyart et al. \cite{Wyart2} further argued for diverging longitudinal and transverse lengths, $\ell_L$ and $\ell_T$, with $\ell_L\sim\ell^*\sim 1/\delta z$ and $\ell_T\sim 1/\sqrt{\delta z}$.  

For the harmonic elastic interaction considered in this work, the pressure above jamming is found \cite{OHern, Wyart2} to scale as $p\sim \delta z^2$, and since the stress per particle $\tilde p =\Gamma/N = pV/N$, we can then write for the scaling of these lengths, 
\begin{equation}
\ell^*\sim\ell_L\sim 1/\tilde p^{1/2},\qquad \ell_T\sim 1/\tilde p^{1/4}.
\end{equation}
If $\ell_L$ (or $\ell_T$) set the length scale for the onset of the anomalously large stress fluctuations reported in this work, then we would expect that, when plotting $C_\Gamma(\mathbf{q})=\langle\Gamma_\mathbf{q}\Gamma_{-\mathbf{q}}\rangle/V$ vs $q\ell_L\sim q/\tilde p^{1/2}$ (or vs $q\ell_T\sim q/\tilde p^{1/4}$), the onset of the anomalous fluctuations at small $q\lesssim q_0$ for different values of $\tilde p$ would all line up at the same value of $q_0/\tilde p^{1/2}$ (or same value of $q_0/\tilde p^{1/4}$).  In Figs.~\ref{Gq-vs-qell}(a), (b) and (c), we therefore plot $C_\Gamma(q\mathbf{\hat y})$ vs $q$, $q/\tilde p^{1/2}$, and $q/\tilde p^{1/4}$ respectively, for the range of $\tilde p=0.00014$ to $0.01831$ (corresponding to the range $\delta z = 0.056$ to $0.75$ \cite{WuHyper}).  We show only data below the peak seen in Fig.~\ref{Gqotildep2-vs-q}, since the high$-q$ data at this peak represent fluctuations on the small length scales of individual particles, which would not be expected to obey any critical scaling.  In Fig.~\ref{Gq-vs-qell} the solid horizontal lines extrapolate through the region where the curves are approximately constant, while the vertical lines denote the approximate point where the curve of $C_\Gamma(q\mathbf{\hat y})$ at the smallest $\tilde p=0.00014$ departs from this horizontal line as $q$ decreases.  These solid lines serve as guides to the eye; if the set of curves were scaling according to the variable on the horizontal axis, we would expect that for all values of $\tilde p$, 
the vertical line would mark the departure of the curve from the  $q-$independent constant represented by the corresponding horizontal line.

Considering Fig.~\ref{Gq-vs-qell}(a), where we plot simply vs $q$, we see that there does appear to be a reasonable alignment of the onset of the small $q$ anomalous fluctuations across all values of $\tilde p$.  The upturn in $C_\Gamma(q\mathbf{\hat y})$ as $q$ decreases seems to take place at roughly the same value of $q_0$ for all $\tilde p$. This is the same conclusion as was previously suggested by Fig.~\ref{Gqotildep2-vs-q}.  In Fig.~\ref{Gq-vs-qell}(b) we see no such alignment at all, thus seemingly ruling out possible scaling with either the isostatic or longitudinal length scales $\ell^*$ and $\ell_L$.  In Fig.~\ref{Gq-vs-qell}(c) the situation is less clear. Looking carefully, one might argue that the curves for the three or four smallest values of $\tilde p$ perhaps do align, with their upturn occurring near the same value of $q/p^{1/4}$; however this is clearly not the case for the larger values of $\tilde p$.  But since scaling is expected to hold only asymptotically close to the jamming transition, i.e. $\tilde p=0$, it could be possible that only these smaller $\tilde p$ are in the proper scaling region.

\begin{figure}
\includegraphics[width=3.2in]{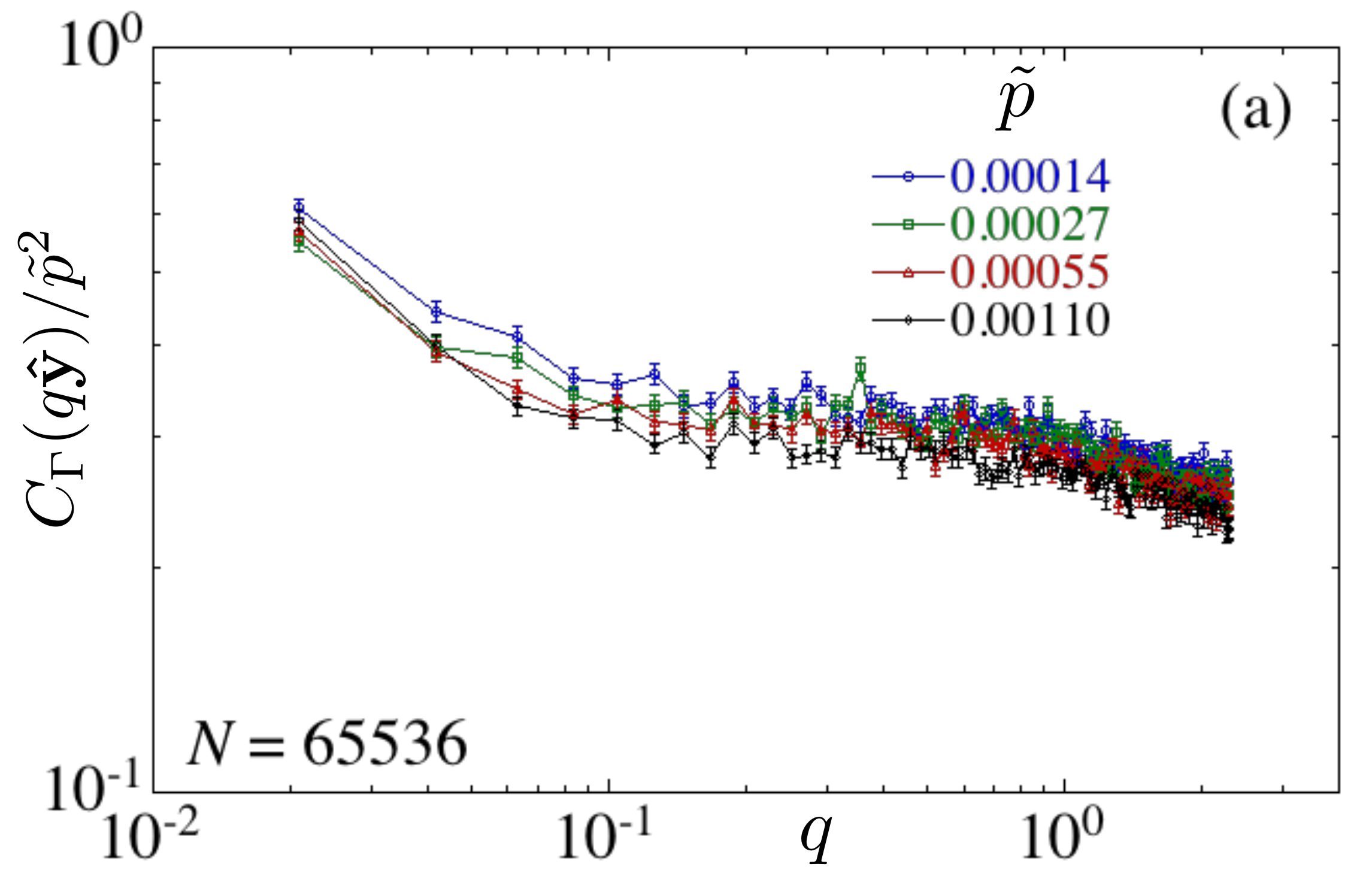}
\includegraphics[width=3.2in]{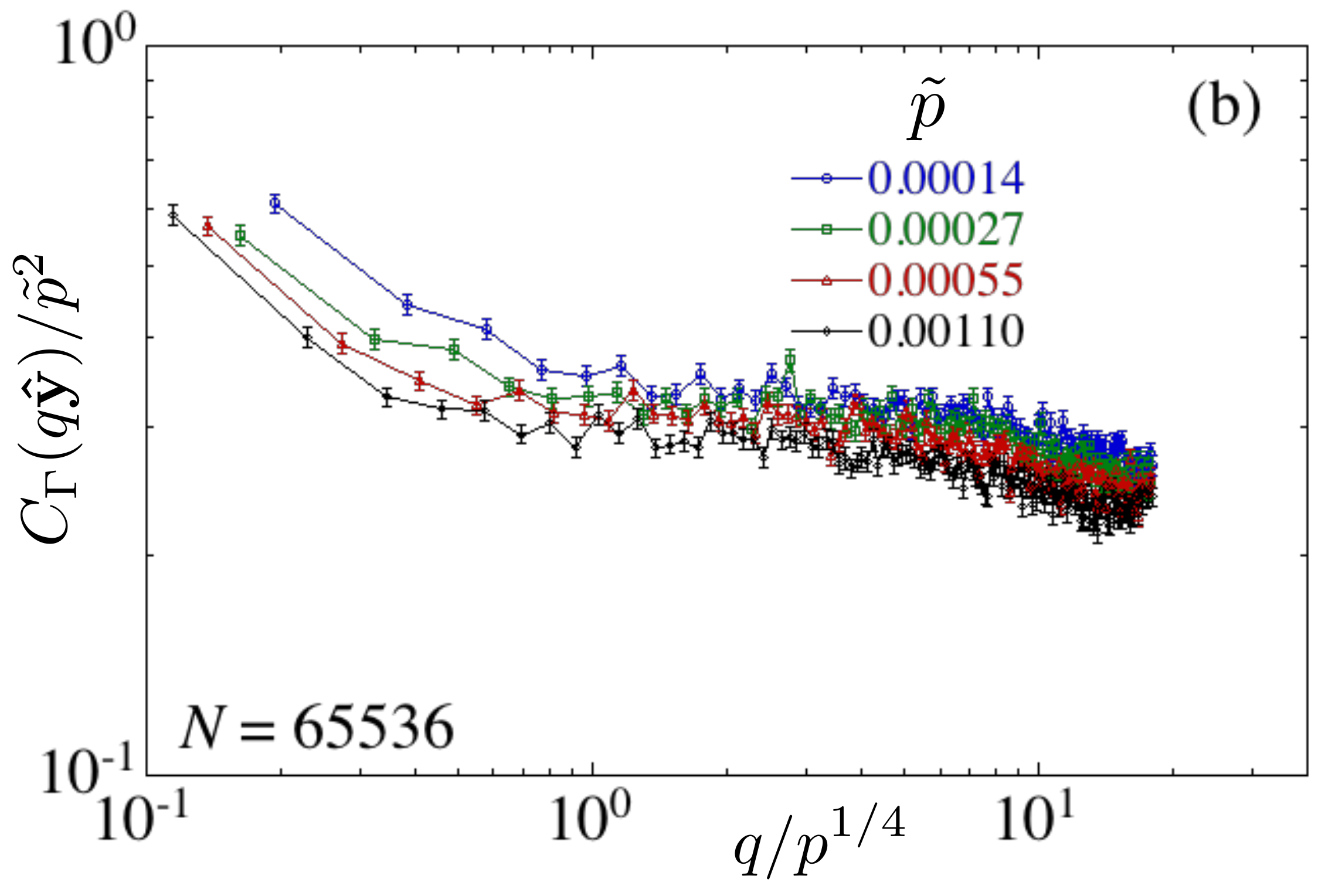}
\caption{(Color online) Plot of $C_\Gamma(q\mathbf{\hat y})/\tilde p^2$ vs (a) $q$ and (b) $q/\tilde p^{1/4}\sim q\ell_T$, for our four smallest values of the total stress per particle $\tilde p=\Gamma/N$ for a system with $N=65536$ particles. Only data for $q$ values below the peak (see Fig.~\ref{Gqotildep2-vs-q}) are shown.  We see that the data collapse looks distinctly better when plotting vs $q$ than vs $q/p^{1/4}$.
}
\label{Gqscale}
\end{figure}

To test for that possibility, we explicitly check whether the curves of $C_\Gamma(q\mathbf{\hat y})$ for these  smallest values of $\tilde p$ can be made to collapse onto each other by rescaling both the horizontal {\em and} vertical axes.  Looking at $C_\Gamma(q_\mathrm{min}\mathbf{\hat y})$ for the smallest value of $q$ in our $N=65536$ size systems, we find that, to excellent agreement, these values scale with the stress per particle as $\tilde p^2$.  In Fig.~\ref{Gqscale} we therefore plot $C_\Gamma(q\mathbf{\hat y})/\tilde p^2$ vs $q$ and vs $q/p^{1/4}$, at our four smallest values of $\tilde p$.  We see that the data collapse looks distinctly better when plotting vs $q$ than when plotting vs $q/p^{1/4}$.  
We thus conclude, from both Figs.~\ref{Gq-vs-qell} and \ref{Gqscale}, that our results are more consistent with $C_\Gamma(q\mathbf{\hat y})/\tilde p^2$ approaching a common limiting curve as $\phi\to\phi_J$ (i.e., as $\tilde p\to 0$), in which the onset of the anomalous fluctuations takes place at a finite value of $q_0$,  than with a $q_0$ that scales to zero as either $1/\ell^*$, $1/\ell_L$ or $1/\ell_T$.

\section{Conclusion}
\label{sconclusions}

To conclude, we find that isotropically compressed, mechanically stable, packings of two dimensional frictionless disks above the jamming transition show anomalously large fluctuations in both isotropic and anisotropic components of the local stress tensor on length scales  larger than  $\ell_0\sim 60$ particle diameters. 
This $\ell_0$ is sufficiently large that earlier numerical studies \cite{Henkes2,Lois,WuTeitel} on smaller systems failed to observe these anomalous fluctuations.  We find that  $\ell_0$ does not appear to vary significantly over the range of pressures studied here, and so there is no evidence that it should be identified with the isostatic length that diverges at jamming \cite{Wyart,Wyart2}.  We have shown that these anomalous stress fluctuations are robust and do not seem to depend on details of the preparation protocol for creating our jammed packings.  

The anomalous stress fluctuations manifest themselves in Fourier space by stress correlations at small wavevectors that increase as $q\to 0$.  This  implies a breakdown of stress self-averaging, as we have directly shown by measuring fluctuations of stress on spatial windows of increasing length $R$.

We find similar anomalous stress fluctuations in the inherent states of a quenched Lennard-Jones liquid, thus leading us to speculate that such fluctuations may be a general feature of amorphous solids in two dimensions.  
The origin of these anomalous fluctuations remains unknown.

\section*{Acknowledgments}

This work was supported by NSF Grant Nos. DMR-1205800 and DMR-1056564.  KK acknowledges financial support from ERC grant ADG20110209.  Computations were carried out in part at the Center for Integrated Research Computing at the University of Rochester.  We thank A.~Lema{\^i}tre for helpful discussions.

\section*{Appendix A}
\label{sA}

To  minimize the energy functional $\tilde U$ of Eq.~(\ref{eUtilde}), and so construct our mechanically stable jammed configurations,
we use the Polak-Ribiere conjugate gradient algorithm \cite{NR} applied to a $2N+3$ dimensional space defined by the $N$ particle positions $\mathbf{r}_i=(x_i,y_i)$ and the three box geometry parameters $L_x$, $L_y$ and $\gamma$ of Fig.~\ref{box}.  Each ``step" of the minimization corresponds to the choice of a new search direction in this $2N+3$ parameter space.
We consider the minimization converged when we satisfy the condition $(\tilde U_i - \tilde U_{i+50})/\tilde U_{i+50} < \varepsilon$, where $\tilde U_i$ is the value at the $i$th step of the minimization and $\varepsilon$ is a suitably small number.  For the results  in the main section of this paper we have used $\varepsilon=10^{-10}$.

\begin{figure}
\includegraphics[width=3.2in]{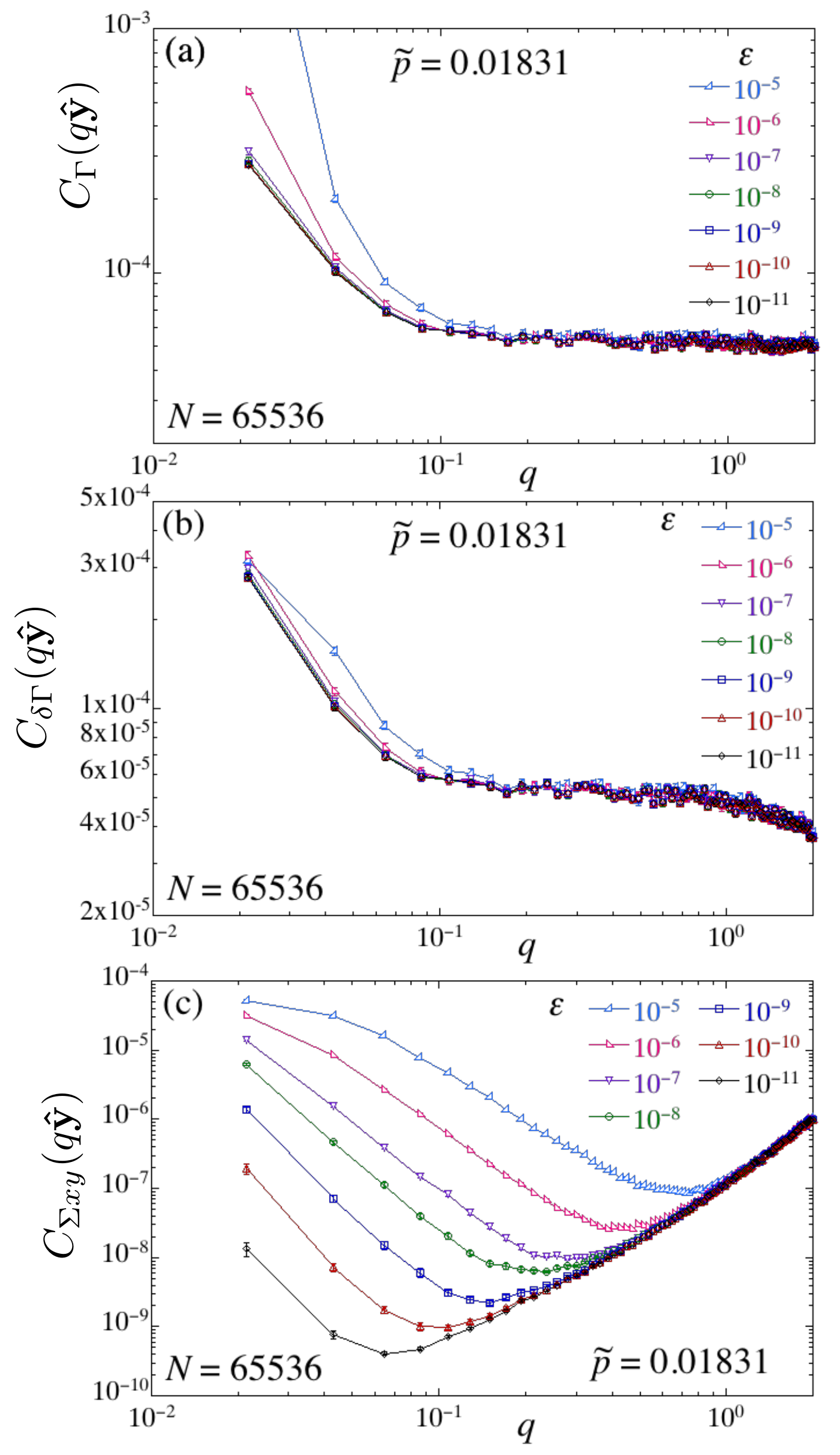}
\caption{(Color online) Stress correlations (a) $C_\Gamma(q\mathbf{\hat y})$, (b) $C_{\delta\Gamma}(q\mathbf{\hat q})$, and (c) $C_{\Sigma_{xy}}(q\mathbf{\hat y})$ vs $q$ for $\tilde p=0.01831$ and $N=65536$ particles.  Results are shown for different values of the minimization convergence parameter $\varepsilon=10^{-5}$ to $10^{-11}$.
}
\label{C-1200}
\end{figure}

\begin{figure}
\includegraphics[width=3.2in]{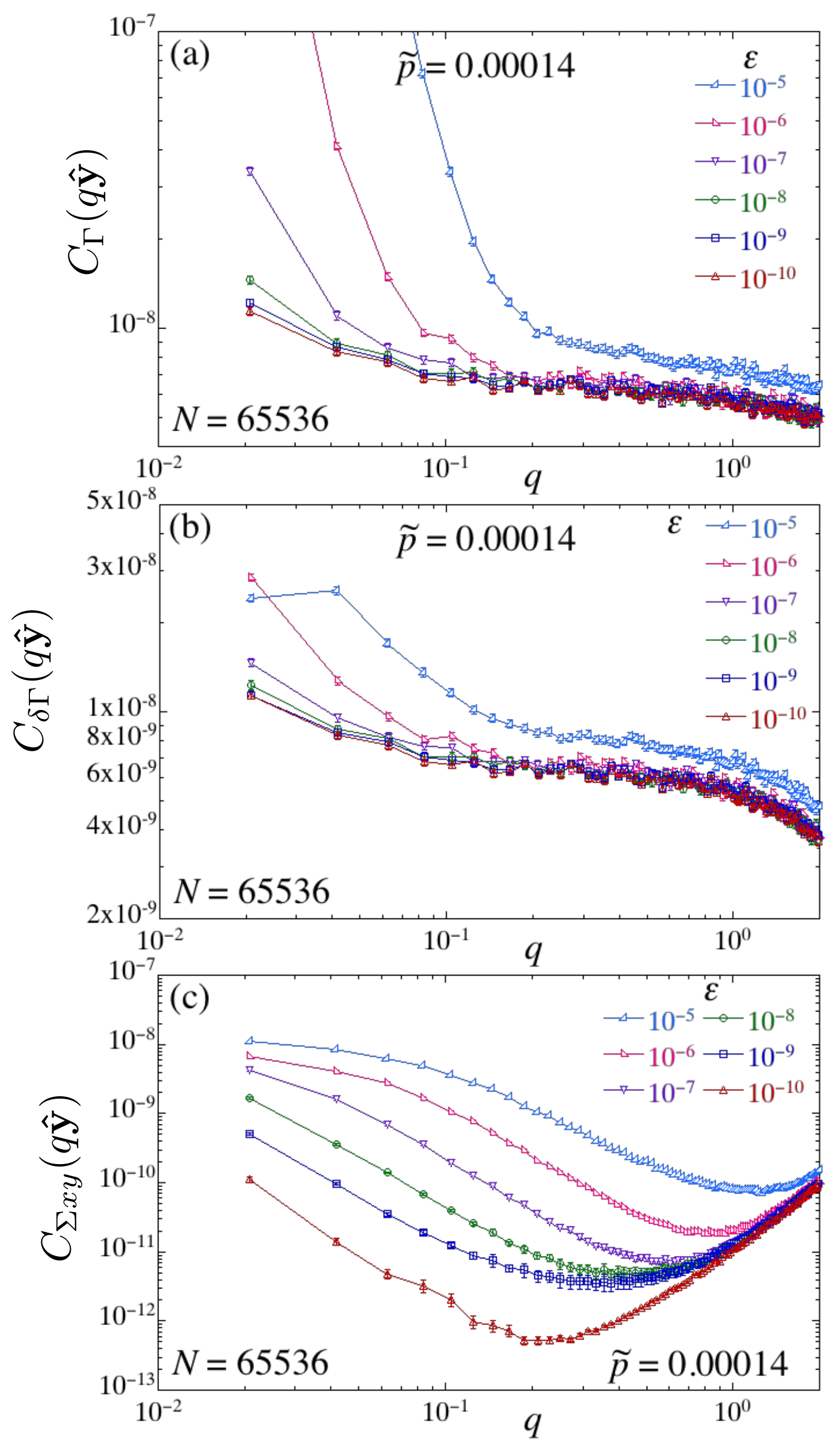}
\caption{(Color online) Stress correlations (a) $C_\Gamma(q\mathbf{\hat y})$, (b) $C_{\delta\Gamma}(q\mathbf{\hat q})$, and (c) $C_{\Sigma_{xy}}(q\mathbf{\hat y})$ vs $q$ for $\tilde p=0.00014$ and $N=65536$ particles.  Results are shown for different values of the minimization convergence parameter $\varepsilon=10^{-5}$ to $10^{-10}$.
}
\label{C-9}
\end{figure}

Tests of how well  this procedure gives configurations with the desired properties have been discussed previously in the Appendix of Ref.~\cite{WuHyper}, which considered the sample to sample fluctuation in the 
box geometry parameters, the accuracy to which the target isotropic total stress tensor is achieved, and the distribution of residual net forces on individual particles in the minimized configurations.
In the present appendix we explicitly test how the stress correlations of Eq.~(\ref{eCq}) behave as we vary the minimization convergence parameter $\varepsilon$.  

In Fig.~\ref{C-1200} we plot the correlations $C_\Gamma(q\mathbf{\hat y})$,  $C_{\delta\Gamma}(q\mathbf{\hat y})$ and $C_{\Sigma{xy}}(q\mathbf{\hat y})$ vs $q$, for a system with $N=65536$ particles and a stress per particle $\tilde p=0.01831$, the largest $\tilde p$ that we consider. Our results are averaged over 1000 independent random initial configurations.  In each case, we show the correlation as it looks when the minimization has been run only up to the convergence parameter $\varepsilon$, which we vary from $\varepsilon=10^{-5}$ to $10^{-11}$.  We see that as $\varepsilon$ decreases, the value of the correlation at small $q$ tends to decrease.  For $C_\Gamma(q\mathbf{\hat y})$ and $C_{\delta\Gamma}(q\mathbf{\hat y})$, shown in Figs.~\ref{C-1200}(a) and (b) respectively, we see that the curves have converged and become independent of $\varepsilon$ once $\varepsilon\le 10^{-8}$.  For $C_{\Sigma_{xy}}(q\mathbf{\hat y})$ in Fig.~\ref{C-1200}(c), however, we do not find convergence even down to our smallest $\varepsilon=10^{-11}$; the value of $C_{\Sigma_{xy}}(q\mathbf{\hat y})$ at small $q$ seems to continually decreases as $\varepsilon$ is made ever smaller.
 We are unable to go to smaller than $\varepsilon=10^{-11}$ due to limitations on our computational ability.  
In Fig.~\ref{C-9} we show the corresponding  correlations at our smallest $\tilde p=0.00014$, where we find similar results.

We thus find that, as $\varepsilon$ decreases,  our constant stress ensemble converges nicely for the correlations $C_\Gamma(q\mathbf{\hat y})$ and  $C_{\delta\Gamma}(q\mathbf{\hat y})$, but has not yet converged for $C_{\Sigma_{xy}}(q\mathbf{\hat y})$.  In order to further examine this latter correlation we  look instead at a constant volume ensemble.  Recall that a comparison between the fixed stress and fixed volume ensembles in Fig.~\ref{compare} showed good agreement for the correlations $C_\Gamma(q\mathbf{\hat y})$ and  $C_{\delta\Gamma}(q\mathbf{\hat y})$.  But the fixed volume ensemble has the advantage that, by keeping the system box fixed, one can get better accuracy in particle force balance, as was found previously in Ref.~\cite{WuHyper} (see Fig.~25 of that work).  We thus compute $C_{\Sigma_{xy}}(q\mathbf{\hat y})$ for a fixed volume ensemble, using the same system parameters as those considered in Fig.~\ref{compare}.  Looking at the packing fraction $\phi=0.88$, corresponding to a relatively high pressure $\langle\tilde p\rangle=0.018$, we find that we are able to achieve force balance to an accuracy of roughly $\max[|\mathbf{F}_i|/\sum_j^\prime |\mathbf{F}_{ij}|]\le 10^{-8}$, where $\mathbf{F}_{ij}$ is the contact force between particle $i$ and $j$, and $\mathbf{F}_i=\sum_j^\prime \mathbf{F}_{ij}$ is the net residual force on particle $i$; the sum is over all particles $j$ in contact with $i$.  This is several orders of magnitude greater accuracy than we were able to achieve in the constant stress ensemble.

\begin{figure}
\includegraphics[width=3.2in]{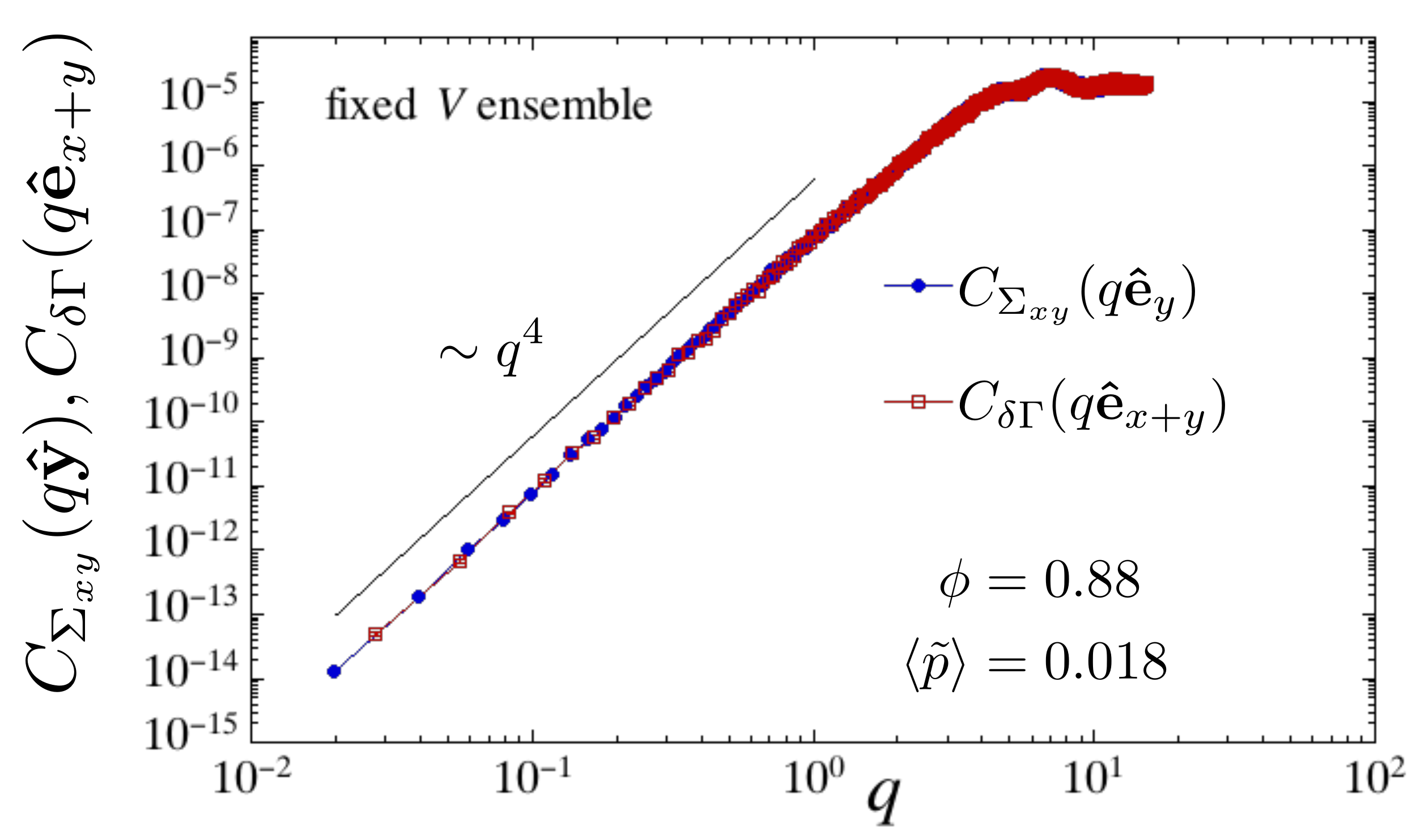}
\caption{(Color online) Stress correlations $C_{\Sigma_{xy}}(q\mathbf{\hat y})$ and $C_{\delta\Gamma}(q\mathbf{\hat e}_{x+y})$ vs $q$ for a constant volume ensemble at fixed packing fraction $\phi=0.88$, corresponding to an average stress per particle $\langle\tilde p\rangle=0.018$. The system box has length $L=320$ and there are $N=77523$ particles.
}
\label{CSxyV}
\end{figure}

In Fig.~\ref{CSxyV} we plot $C_{\Sigma_{xy}}(q\mathbf{\hat y})$ vs $q$ for this constant volume ensemble.  For comparison, we also plot $C_{\delta\Gamma}(q\mathbf{\hat e}_{x+y})$ for this same constant volume ensemble.  Assuming  the rotational isotropy of fluctuations as in Eq.~(\ref{rotate2}), these two correlations should be equal.  We see that these correlations are indeed equal, and that they go algebraically to zero as $q$ vanishes.  Fitting to the linear part of the curve on the log-log plot, we find $C_{\Sigma_{xy}}(q\mathbf{\hat y})=C_{\delta\Gamma}(q\mathbf{\hat e}_{x+y})\sim q^4$. 
Thus the HC result of Eq.~(\ref{sigHC}),  which predicts that this correlation should vanish at all $q$, is found to hold only in the $q\to 0 $ limit.

\section*{Appendix B}
\label{sB}

Here we derive Eq.~(\ref{eDC}) relating the real space fluctuations $\Delta_X(R)$ to the correlations $C_X(\mathbf{q})$.  We will give our derivation in terms of the isotropic part of the stress $\Gamma$, but the same arguments hold for $\delta\Gamma$ and $\Sigma_{xy}$.  

We define a local pressure field $p(\mathbf{r})$.  For our calculations in Sec.~\ref{ssoftR} we have used,
\begin{equation}p(\mathbf{r})=\sum_i\Gamma_i\delta(\mathbf{r}-\mathbf{r}_i),
\label{epr} 
\end{equation}
but one could instead use a coarse grained function.  The total stress on a circular window of radius $R$ is defined as,
\begin{equation}
\Gamma_R=\int_R d^2r\,p(\mathbf{r}),
\end{equation}
where the integral is over a circle of radius $R$.  We then define the Fourier transforms,
\begin{equation}
\Gamma_\mathbf{q}=\int_V d^2r\,\mathrm{e}^{i\mathbf{q}\cdot\mathbf{r}}p(\mathbf{r}),\quad
p(\mathbf{r})=\dfrac{1}{V}\sum_\mathbf{q}\mathrm{e}^{-i\mathbf{q}\cdot\mathbf{r}}\,\Gamma_\mathbf{q},
\end{equation}
where the integral is over the entire system of volume $V$, and the sum is over all allowed wavevectors given by Eq.~(\ref{eqs}).  Note, $\Gamma_{\mathbf{q}=0}=\int_Vd^2r\,p(\mathbf{r})=\Gamma$, the total stress on the system.

We then have, 
\begin{equation}
\begin{aligned}
\langle\Gamma_R^2\rangle &= \int_R d^2r\int_R d^2r^\prime\,\langle p(\mathbf{r})p(\mathbf{r}^\prime)\rangle\\
&=\int_Rd^2r\int_Rd^2r^\prime\,\dfrac{1}{V^2}\sum_{\mathbf{q},\mathbf{q}^\prime}
\mathrm{e}^{-i\mathbf{q}\cdot\mathbf{r}}\mathrm{e}^{-i\mathbf{q}^\prime\cdot\mathbf{r}^\prime}
\langle\Gamma_\mathbf{q}\Gamma_{\mathbf{q}^\prime}\rangle.
\end{aligned}
\end{equation}
Assuming the ensemble averaged pressure correlations have translational invariance, i.e.,
\begin{equation}\langle p(\mathbf{r})p(\mathbf{r}^\prime)\rangle=\langle p(\mathbf{r}-\mathbf{r}^\prime)p(0)\rangle, 
\end{equation}
we have 
\begin{equation}
\langle\Gamma_\mathbf{q}\Gamma_{\mathbf{q}^\prime}\rangle=\delta_{-\mathbf{q},\mathbf{q}^\prime}\langle\Gamma_\mathbf{q}\Gamma_{-\mathbf{q}}\rangle,
\end{equation}
and the above becomes,
\begin{equation}
\begin{aligned}
\langle\Gamma_R^2\rangle&=\int_Rd^2r\int_Rd^2r^\prime\,\dfrac{1}{V}\sum_\mathbf{q}\mathrm{e}^{-i\mathbf{q}\cdot(\mathbf{r}-\mathbf{r}^\prime)}C_\Gamma(\mathbf{q})\\
&=\dfrac{1}{V}\sum_\mathbf{q}C_\Gamma(\mathbf{q})\left[\int_Rd^2r\,\mathrm{e}^{-i\mathbf{q}\cdot\mathbf{r}}\right]\left[\int_Rd^2r^\prime\,\mathrm{e}^{i\mathbf{q}\cdot\mathbf{r}^\prime}\right].
\end{aligned}
\end{equation}
Each of the terms in the square brackets above is just the Fourier transform of the indicator function $D(\mathbf{r})$ for a circle of radius $R$, i.e. $D(\mathbf{r})=1$ for $\mathbf{r}$ within the circle, and $D(\mathbf{r})=0$ otherwise,
\begin{equation}
D_\mathbf{q}=\int_Rd^2r\,\mathrm{e}^{-i\mathbf{q}\cdot\mathbf{r}}=\pi R^2f(|\mathbf{q}|R),
\end{equation}
with,
\begin{equation}
f(u)=\dfrac{2}{u^2}\int_0^udv\,vJ_0(v),
\end{equation}
and $J_0(v)$ the Bessel function of the first kind.  Thus,
\begin{equation}
\langle\Gamma_R^2\rangle=\dfrac{(\pi R^2)^2}{V}\sum_\mathbf{q}C_\Gamma(\mathbf{q})f^2(qR).
\label{eDC1}
\end{equation}
Next, noting that $\langle\Gamma_R\rangle/(\pi R^2)=\Gamma/V$, we have,
\begin{equation}
\langle\Gamma_R\rangle^2=\dfrac{(\pi R^2)^2}{V^2}\Gamma^2=\dfrac{(\pi R^2)^2}{V}C_\Gamma(0).
\label{eDC2}
\end{equation}
Finally, noting that $f(0)=1$, Eq.~(\ref{eDC2}) is just the $\mathbf{q}=0$ term of Eq.~(\ref{eDC1}), and we thus get Eq.~(\ref{eDC}),
\begin{equation}
\Delta_\Gamma(R)=\dfrac{\langle\Gamma_R^2\rangle-\langle\Gamma_R\rangle^2}{\pi R^2}=\dfrac{\pi R^2}{V}\sum_{\mathbf{q}\ne 0}C_\Gamma(\mathbf{q})f^2(qR).
\end{equation}

Note, the real space fluctuation measure $\Delta_\Gamma(R)$ involves a sum on $C_\Gamma(\mathbf{q})$ over {\em all} $\mathbf{q}$.  While we expect that $C_\Gamma(\mathbf{q})$ at {\em small} $\mathbf{q}$ is independent of the details of how $p(\mathbf{r})$ is defined on {\em short} length scales, i.e. whether we use our $p(\mathbf{r})$ given by Eq.~(\ref{epr}) or whether we use a coarse grained version, the correlation $C_\Gamma(\mathbf{q})$ does depend on such details at large $\mathbf{q}$.  Depending on the size of the system, and the size of the window $R$, the small length scale behavior of $p(\mathbf{r})$ can significantly affect the observed value of $\Delta_R(\Gamma)$, as has been reported recently \cite{WuHyper} for the corresponding fluctuations of the local packing fraction.  Only in the limit of sufficiently large $R$ will $\Delta_\Gamma(R)$ become independent of the small length scale behavior of $p(\mathbf{r})$.  The results reported in Sec.~\ref{ssoftR} are thus only for the specific choice of $p(\mathbf{r})$ given in Eq.~(\ref{epr}).


\begin{thebibliography}{99}
\bibitem{LJnote}Our LJ potential is cut off at a distance $r=2.5d_{ij}$ using lj/cut in LAMMPS.
\end{thebibliography}
\end{document}